 \documentclass[prd,showpacs]{revtex4}% Physical Review D

 \usepackage[dvips,final]{graphicx}
  \usepackage{amssymb}
   \usepackage{amsmath}
    \usepackage{amsfonts}
     \usepackage{epsfig}
      \usepackage{bm}% bold math
\usepackage{graphicx}% Include figure files
\usepackage{dcolumn}% Align table columns on decimal point
\usepackage{bm}% bold math
\def\a {\epsilon}
\def\g {\gamma}
\def\M  {{\cal M}}
\def\F  {{\cal F}}
\def\O  {{\cal O}}
\def\p  {{\cal P}}
\def\R  {{\cal R}}
\def\J {$J/\psi$ }
\def\j { J/\psi  }

\def\cc {$c\bar{c}$ }
\def\cj {$\chi_{cJ}$ }

\def\ktf {$k_t$-factorization }
\def\ktfa {$k_t$-factorization approach }
\def\cpc#1#2#3  {{Computer\ Phys.\ Comm.\ }  {\bf#1}, #2 (#3)}
\def\err#1#2#3  {{\it Erratum }              {\bf#1}, #2 (#3)}
\def\epjc#1#2#3 {{Eur. Phys. J. C }          {\bf#1}, #2 (#3)}
\def\dum#1#2#3  {{~}                         {\bf#1}, #2 (#3)}
\def\ib#1#2#3   {{\it ibid. }                {\bf#1}, #2 (#3)}
\def\jcp#1#2#3  {{J.\ Comput.\ Phys.\ }      {\bf#1}, #2 (#3)}
\def\jhep#1#2#3 {{JHEP }                     {\bf#1}, #2 (#3)}
\def\ijmp#1#2#3 {{Int.\ J.\ Mod.\ Phys.\ }   {\bf#1}, #2 (#3)}
\def\jpg#1#2#3  {{J.\ Phys.\ G }             {\bf#1}, #2 (#3)}
\def\mpl#1#2#3  {{Mod.\ Phys.\ Lett.\ }      {\bf#1}, #2 (#3)}
\def\ncim#1#2#3 {{Nuovo Cimento }            {\bf#1}, #2 (#3)}
\def\np#1#2#3   {{Nucl.\ Phys.\ }            {\bf#1}, #2 (#3)}
\def\pan#1#2#3  {{Phys.\ At.\ Nuclei }       {\bf#1}, #2 (#3)}
\def\plb#1#2#3  {{Phys.\ Lett.\ B }          {\bf#1}, #2 (#3)}
\def\prep#1#2#3 {{Phys.\ Rep.\ }             {\bf#1}, #2 (#3)}
\def\prd#1#2#3  {{Phys.\ Rev.\ D }           {\bf#1}, #2 (#3)}
\def\prl#1#2#3  {{Phys.\ Rev.\ Lett.\ }      {\bf#1}, #2 (#3)}
\def\ptp#1#2#3  {{Prog.\ Theor.\ Phys.\ }    {\bf#1}, #2 (#3)}
\def\ps#1#2#3   {{Physica Scripta }          {\bf#1}, #2 (#3)}
\def\rmp#1#2#3  {{Rev.\ Mod.\ Phys.\ }       {\bf#1}, #2 (#3)}
\def\rpp#1#2#3  {{Rep.\ Prog.\ Phys.\ }      {\bf#1}, #2 (#3)}
\def\sa#1#2#3   {{Sci. Acta}                 {\bf#1}, #2 (#3)}
\def\sjnp#1#2#3 {{Sov.\ J.\ Nucl.\ Phys.\ }  {\bf#1}, #2 (#3)}
\def\spj#1#2#3  {{Sov.\ Phys.\ JETP }        {\bf#1}, #2 (#3)}
\def\spjl#1#2#3 {{Sov.\ JETP Lett.\ }        {\bf#1}, #2 (#3)}
\def\spu#1#2#3  {{Sov.\ Phys.-Usp.\ }        {\bf#1}, #2 (#3)}
\def\yaf#1#2#3  {{Yad.\ Fiz.\ }              {\bf#1}, #2 (#3)}
\def\zp#1#2#3   {{Zeit.\ Phys.\ }            {\bf#1}, #2 (#3)}
\def\zpc#1#2#3  {{Z.\ Phys.\ C }             {\bf#1}, #2 (#3)}
\def\etal {{\it et al. }}
\begin{document}

\thispagestyle{empty} \preprint{\hbox{}} \vspace*{-10mm}

\title{
%Learning the dynamics of gluon ladders at intermediate $x$\\
Inclusive production of $J/\psi$ meson \\
in proton-proton collisions at the BNL RHIC}
\author{S.\ P.\ Baranov}
\email{baranov@sci.lebedev.ru}
\affiliation{P.N. Lebedev Institute of Physics,
             Lenin avenue 53, Moscow 119991, Russia}
%\marginpar{$\surd$ 1}
%
\author{A.\ Szczurek}
\email{antoni.szczurek@ifj.edu.pl}
\affiliation{Institute of Nuclear Physics PAN, PL-31-342 Cracow,
Poland} \affiliation{University of Rzesz\'ow, PL-35-959 Rzesz\'ow,
Poland}
\date{\today}

\begin{abstract}
Inclusive cross sections for $J/\psi$ production
in proton-proton collisions were calculated in the $k_t$-factorization
approach for the RHIC energy. Several mechanisms were considered,
including direct color-singlet mechanism, radiative decays of $\chi_c$
mesons, decays of $\psi'$, open-charm associated production of $J/\psi$
as well as weak decays of B mesons. Different unintegrated gluon
distributions from the literature were used.
We find that radiative $\chi_c$ decays and direct color-singlet
contributions constitute the dominant mechanism of $J/\psi$ production.
These process cannot be consistently treated within
collinear-factorization approach.
The results are compared with recent RHIC data.
The new precise data at small transverse momenta impose stringent
constraints on UGDFs. Some UGDFs are inconsistent with the new data.
The Kwieci\'nski UGDFs give the best description of the data.
In order to verify the mechanism suggested
here we propose $J/\psi$ -- jet correlation measurement and
an independent measurement of $\chi_c$ meson production
in $\pi^+ \pi^-$ and/or $K^+ K^-$ decay channels.
Finally, we address the issue of \J spin alignment.  
%\marginpar{$\surd$ 2}
\end{abstract}

%Keywords:
\maketitle

%-----------------------------
\section{Introduction}
%-----------------------------

For the last decade, the inclusive production of \J mesons was a serious
theoretical puzzle challenging our understanding of QCD, parton model,
and the bound state formation dynamics.
The roots of the puzzling \J history trace back to the middle 1990s, when 
the data on \J and $\Upsilon$ hadroproduction cross sections 
\cite{CDF1}-\cite{CDF3}
revealed a more than one order-of-magnitude discrepancy with theoretical 
expectations. This fact has induced extensive theoretical activity and led 
to the introduction of new production mechanisms, known as the color-octet 
model \cite{BBL,ChoLei} and gluon vector dominance model \cite{GVDM}.
Since then, the color-octet model has been believed to give the most 
likely explanation of the quarkonium production phenomena, although there 
were also some indications that it was not working well. The situation 
became even more intriguing after the measurements of \J spin alignment 
\cite{E537}-\cite{CDF4} have been carried out showing inconsistency
with the newly accepted theory.
% and the experiment, once again, has demonstrated 
% dramatic disagreement with the newly accepted and commonly trusted theory.

At the same time, it has been shown that the incorporation of the usual
color-singlet production scheme with the \ktfa can provide a reasonable
and consistent picture of the phenomenon under study in its entirety.
Whithin the latter approach, a good description of data on the production 
of $J/\psi$, $\chi_c$, and $\Upsilon$ mesons both at the Tevatron 
\cite{J_Tev,HKSST01} and HERA \cite{J_HERA} has been achieved, and even 
a solution to the \J spin alignment problem has been guessed 
\cite{J_Tev,J_spin}. 
The issue of the quarkonium production mechanism continues 
to be under intense debate.

Recently, the PHENIX collaboration at the 
BNL Relativistic Heavy Ion Collider (RHIC) has measured inclusive \J
production in elementary proton-proton collisions \cite{RHIC_JPSI}.
While for the RHIC community the elementary $pp$ cross section is only
the baseline for the nuclear case, we wish to demonstrate that
the elementary data by itself constitute a very valueable information
about QCD dynamics in the region of intermediate 
$x\simeq$ 10$^{-2}$--10$^{-1}$. 
In our paper we present a detailed analysis of RHIC data based on
the \ktfa and a large variety of unintegrated gluon distribution
functions (UGDFs). We show that the
new precise data at small \J transverse momenta impose stringent
constraints on UGDFs and, consequently, stimulate better understanding
of the underlying gluon dynamics. 

The outline of the paper is the following. In Sec. II we describe the 
production mechanisms employed in our analysis and discuss the different 
parametrizations of UGDFs. In Sec. III we compare our theoretical 
predictions with experimental results and derive new predictions on 
the quantities which at yet have not been measured but could serve as
important cross-check of our understanding of the reaction mechanism.
Our findings and recommendations for the forthcoming experiments are
summarized in Sec. IV.

%---------------------------
\section{Formalism}
%---------------------------

%--------------------------------------------------------------------
\subsection{Different mechanisms of $J/\psi$ production}
%--------------------------------------------------------------------
%\subsection{Production mechanisms}

In this paper, we take into account a number of different mechanisms
leading to the appearance of \J mesons in the final state (of course,
they are not thought to be all of equal importance). The considered
mechanisms are the following.

Direct color-singlet \J production via gluon-gluon fusion
\begin{equation}\label{direct}
g+g\to J/\psi+g;
\end{equation}
direct production of $\psi'$ meson
\begin{equation}\label{prime}
g+g\to\psi'+g
\end{equation}
and its subsequent decay $\psi'\to J/\psi+X$;\\
production of $P$-wave charmonium states $\chi_{cJ}\;$ $(J=0,1,2)$
\begin{equation}\label{chi}
g+g\to\chi_{cJ}
\end{equation}
followed by their radiative decays $\chi_{cJ}\to J/\psi+\gamma$;\\
production of $b$ quarks and antiquarks
\begin{equation}\label{bb}
g+g\to b+\bar{b}
\end{equation}
followed by their fragmentation into $B$ mesons and subsequent weak
decays $B\to J/\psi+X$;\\
production of \J mesons in association with unbound charmed quarks
\begin{equation}\label{jcc}
g+g\to J/\psi+c+\bar{c}.
\end{equation}
%production of color-octet states 
%\begin{equation}\label{com}
%g+g\to c\bar{c}[^{2S{+}1}L_{J}^{(8)}]+g
%\end{equation}
%followed by nonperturbative transitions into \J mesons.
Examples of the relevant Feynman diagrams for all the mentioned 
processes are shown in Fig. 1. Every subprocess is accompanied by the
emission of gluon jets, as is shown in Fig. 2.    

In general, there could also exist color-octet contributions.
The latter cannot be calculated from the first principles and are 
usually estimated from fits to existing data. It has been shown already
that, within the \ktf approach, these contribuions are consistent with 
zero both at the Tevatron \cite{J_Tev} and HERA \cite{J_HERA}.
In view of the uncertainties coming from other contributions we find it
not useful to include color-octet contributions in the present analysis.

%The color octet contributions (diagrams (e)) cannot be calculated
%from first principles. Usually it is fitted to existing
%data. In principle, it could be fitted including
%other contributions that can be calculated from
%first principles. However, in the light of uncertainties
%in calculating the other contributions we do not find
%such a procedure attractive and useful.

%\marginpar{$\surd$ 3}
%------------------------------------------------------------
%\begin{figure}[!h]    % 
% {\includegraphics[width=0.6\textwidth]{baranov_fig1.eps}}
%   \caption{\label{fig:diagrams_of_all_processes}
%\small
%Diagrams of processes included in this work.
%}
%\end{figure}
%-------------------------------------------------------------

%------------------------------------------------------------
\begin{figure}[htb]    %  
 {\includegraphics[width=0.6\textwidth]{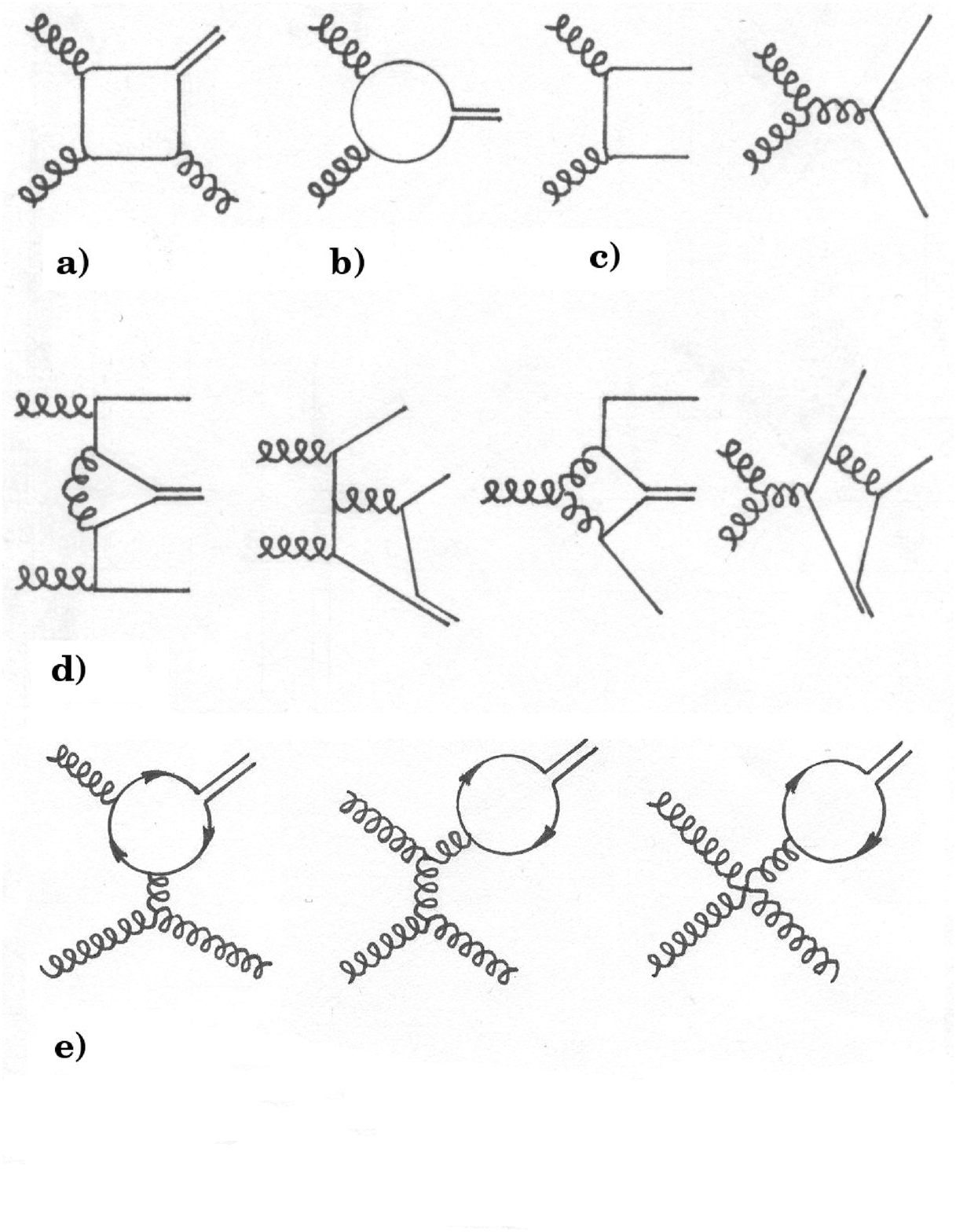}}
   \caption{\label{fig:diagrams_of_all_processes}
\small
Processes included in our approach:
(a) direct color-singlet production, (b) production of $\chi_c$ mesons,
(c) open bottom quark production, (d) open-charm associated production,
(e) color-octet production
}
\end{figure}
%-------------------------------------------------------------

%------------------------------------------------------------
\begin{figure}[htb]    % 
 {\includegraphics[width=0.6\textwidth]{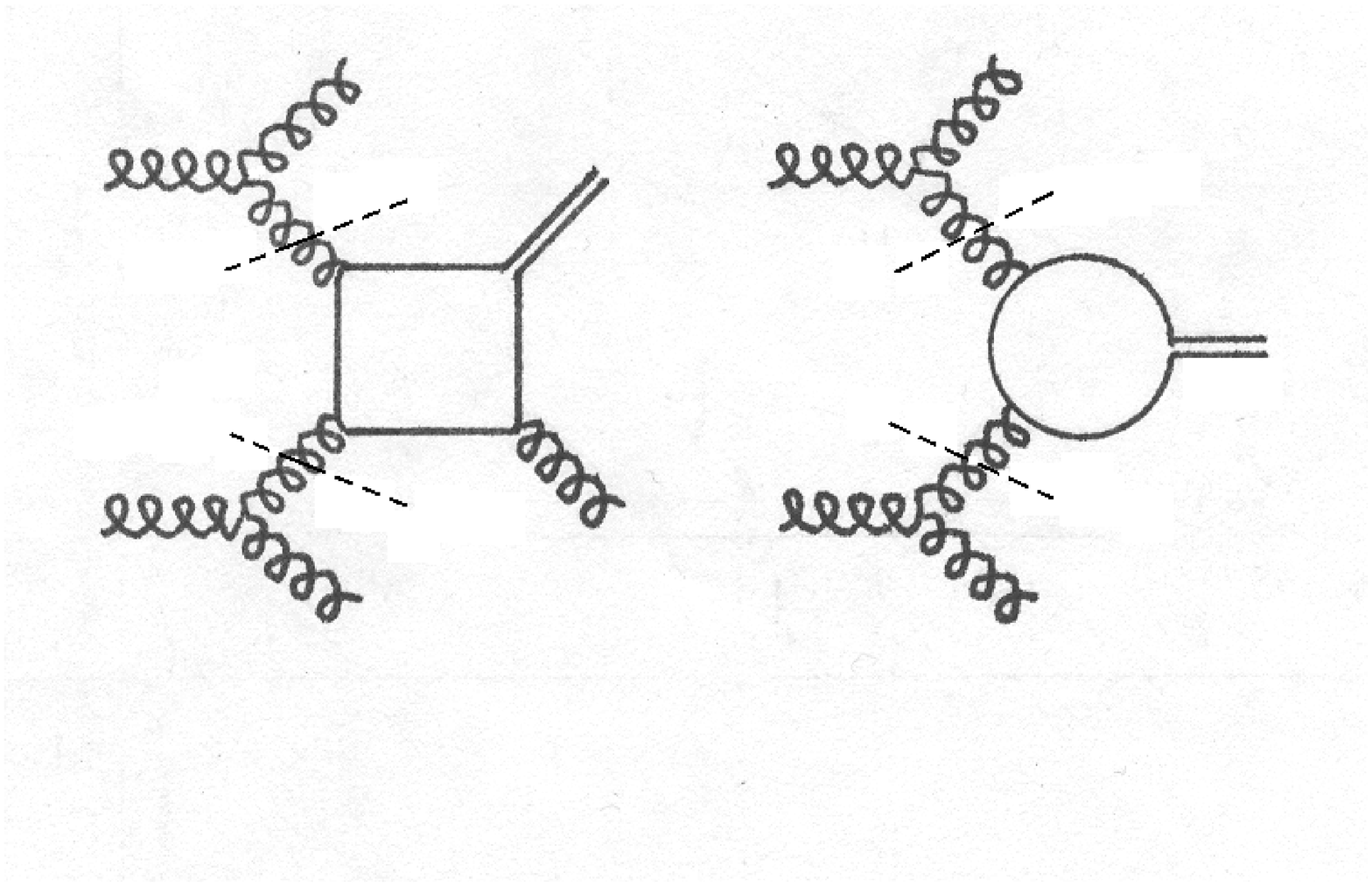}}
   \caption{\label{fig:diagrams_kt_factorization}
\small
Application of UGDFs to inclusive production of $J/\psi$ (left)
and $\chi_c$ (right). The upper and the lower parts of these diagrams 
are included in the $k_t$ evolution of gluon densities. The emitted 
gluons can realise in the final state hadronic jets.
}
\end{figure}
%-------------------------------------------------------------

A few words are in order to describe the formation of \cc bound states.
First of all, it should be noted that the amplitudes of the subprocesses
(\ref{direct})-(\ref{chi}), (\ref{jcc}) contain projection
operators $J(S,L)$, which guarantee the proper quantum numbers of the
\cc state under consideration. These operators read for the different
spin and orbital angular momentum states \cite{BaiBer,GubKra}:
\begin{eqnarray}
\label{j00} J(^1S_0)\equiv
J(S{=}0,L{=}0)&=&\gamma_5\,(\not{p_c}+m_c)/m_{\psi}^{1/2} \; , \\
\label{j10} J(^3S_1)\equiv
J(S{=}1,L{=}0)&=&\not{\epsilon(S_z)}\,(\not{p_c}+m_c)/m_{\psi}^{1/2}
\; , \\
\label{j11} J(^3P_J)\equiv
J(S{=}1,L{=}1)&=&(\not{p_{\bar{c}}}-m_c)\,
                 \not{\epsilon(S_z)}\,(\not{p_c}+m_c)/m_{\psi}^{3/2}
\; ,
\end{eqnarray}
where $m_\psi$ is the mass of the specifically considered \cc state and 
$m_c=m_\psi/2$ the mass of the charmed quark (always set equal to $1/2$ 
of the meson mass, as is required by the nonrelativistic bound-state
model).

States with various projections of the spin momentum onto the $z$ axis 
are represented by the polarization vector $\epsilon(S_z)$.

The probability for the two quarks to form a meson depends on the bound
state wave function $\Psi(q)$. In the nonrelativistic approximation which 
we are using here, the relative momentum $q$ of the quarks in the bound state 
is treated as a small quantity. So, it is useful to represent the quark 
momenta as $ p_{c}=p_{\psi}/2+q,\;\; p_{\bar{c}}=p_{\psi}/2-q$.
Then, we multiply the matrix elements by $\Psi(q)$ and perform 
integration with respect to $q$. The integration is performed after 
expanding the integrand around $q=0$:
\begin{equation}
{\cal M}(q)={\cal M}|_{q=0}+
(\partial{\cal M}/\partial
q^\alpha)|_{q=0}q^\alpha\,+\,\,\dots\label{exp}
\; .
\end{equation}
Since the expressions for
 ${\cal M}|_{q=0}$, $(\partial{\cal M}/\partial q^\alpha)|_{q=0}$, etc.
are no longer dependent on $q$, they may be factored outside the integral
sign. A term-by-term integration of this series then yields
\cite{GubKra}:
\begin{eqnarray}
 &&\int \frac{d^3q}{(2\pi)^3}\,\Psi(q)=\frac{1}{\sqrt{4\pi}}\,\R(x=0),
 \label{r0} \\
 &&\int \frac{d^3q}{(2\pi)^3}\,q^\alpha\Psi(q)=
 -i\epsilon^\alpha(L_z)\,\frac{\sqrt{3}}{\sqrt{4\pi}}\,\R'(x=0),
 \label{r1}
\end{eqnarray}
etc., where $\R(x)$ is the radial wave function in the
coordinate representation (the Fourier transform of $\Psi(q)$). The first
term contributes only to $S$-waves, but vanishes for $P$-waves because
$\R_P(0)=0$. On the contrary, the second term contributes only to
$P$-waves, but vanishes for $S$-waves because $\R'_S(0)=0$.
States with various projections of the orbital angular momentum onto the
$z$ axis are represented by the polarization vector $\epsilon(L_z)$.
The numerical values of the wave functions are either known from the
leptonic decay widths (for \J and $\psi'$ mesons) or can be taken from
potential models (for $\chi_{cJ}$ mesons).
Including radiative corrections changes the values of the wave functions
by a factor of 2 (the NLO result compared to the LO result), and so, one
can also expect large effect from higher order corrections.
This leads to a sizeable theoretical uncertainty, which, on the other 
hand, can only affect the absolute normalization but not the shape of the 
$p_t$ spectrum.

When calculating the spin average of the matrix elements squared, 
we adopt the \ktf prescription \cite{GLR83} for the off-shell gluon spin 
density matrix:
\begin{equation} \label{epskt}
 \overline{\a^{\mu}\a^{*\nu}} = k_{t}^\mu k_{t}^\nu/|k_{t}|^2,
\end{equation}
where $k_{t}$ is the component of the gluon momentum perpendicular to 
the beam axis, and the bar stands for the averaging over the gluon spin.
In the collinear limit, when $k_{t}\to 0$, this expression converges to
the ordinary $\overline{\a_g^{\mu}\a_g^{*\nu}}=-\frac{1}{2}\;g^{\mu\nu}$.
In all other respects, the evaluation of the diagrams is straightforward
and follows the standard QCD Feynman rules. This has been done using the
algebraic manipulation systems FORM \cite{FORM} and REDUCE \cite{REDUCE}.

For the direct production mechanism (\ref{direct}) the fully differential
cross section reads
\begin{eqnarray}
&& d\sigma(pp\to\psi X)=
\frac{\pi\alpha_s^3\,|\R(0)|^2}{\hat{s}^2}\,
\frac{1}{4}\sum_{\mbox{{\tiny spins}}}\,
\frac{1}{64}\sum_{\mbox{{\tiny colors}}}
|{\cal M}(gg\to\psi g)|^2  \nonumber \\ && \times
 \F_g(x_1,k_{1t}^2,\mu^2)\,\F_g(x_2,k_{2t}^2,\mu^2)\,\, \label{ME_direct}
dk_{1t}^2\, dk_{2t}^2\, dp_{\psi T}^2\, dy_3\,dy_{\psi}\,
\frac{d\phi_1}{2\pi}\,\frac{d\phi_2}{2\pi}\, \frac{d\phi_{\psi}}{2\pi}
\; ,
\end{eqnarray}
where $\phi_1$ and $\phi_2$ are the azimuthal angles of the initial 
gluons, and $y_{\psi}$ and $\phi_{\psi}$ the rapidity and the azimuthal 
angle of \J particle. 
The explicit expressions for the parton level matrix elements 
$|{\M}(gg\to \psi g)|^2$ are presented in Ref. \cite{J_Tev}.

The phase space physical boundary is determined by the inequality 
\cite{BycKaj}
\begin{equation}
G(\hat{s}, \hat{t}, k_3^2, k_1^2, k_2^2, m_{\psi}^2) \le 0,
\end{equation}
with $k_1$, $k_2$ and $k_3$ being the initial and final gluon momenta, 
$\hat{s}=(k_1+k_2)^2$, $\;\hat{t}=(k_1-p_{\psi})^2$, and $G$ is the
standard kinematic function \cite{BycKaj}.
The initial gluon momentum fractions $x_1$ and $x_2$ appearing in the 
unintegrated gluon distribution functions $\F_g(x_i,k_{i,t}^2,\mu^2)$ 
are calculated from the energy-momentum conservation laws in the light 
cone projections:
\begin{eqnarray}
 (k_1+k_2)_{E+p_{||}}=
 x_1\sqrt{s} &=& m_{\psi T}\exp(y_{\psi})\;\; + \; |k_{3t}|\exp(y_3),
 \nonumber \\[-2mm] && \label{x1x2} \\[-2mm]
 (k_1+k_2)_{E-p_{||}}=
 x_2\sqrt{s} &=& m_{\psi T}\exp(-y_{\psi}) + |k_{3t}|\exp(-y_3),\nonumber
\end{eqnarray}
where $m_{\psi T}=(m_{\psi}^2+|p_{\psi T}|^2)^{1/2}$.

The production scheme of $\psi'$ meson (\ref{prime}) is identical to that 
of \J, and only the numerical value of the wave function $|\R(0)|^2$ is
different. In both cases, the values of the wave functions were extracted 
from the known leptonic decay widths \cite{PDG} using the formula
$|\R(0)|^2=\Gamma_{ee}m_{\psi}^2/(4\alpha^2e_c^2)\,[1-16\alpha_s/(3\pi)]$
and were set equal to 
$|\R_{J/\psi}(0)|^2 = 0.8$ GeV$^3$ for \J meson, and
$|\R_{\psi'}(0)|^2 = 0.4$ GeV$^3$ for $\psi'$ meson.
To calculate the feed-down to \J states, the $\psi'$ production cross
section has to be multiplied by the branching fraction
$Br(\psi'\to J/\psi X)=56\%$ \cite{PDG}.

For the production of \cj mesons via the subprocess (\ref{chi}) we have
%\marginpar{$\surd$ 4}
%
\begin{eqnarray}
&& d\sigma(pp\to\chi_{cJ}X)=
\frac{12\pi^2\alpha_s^2\,|\R'(0)|^2}%
{x_1x_2s\;\lambda^{1/2}(\hat{s},k_1^2,k_2^2)}\,
\frac{1}{4}\sum_{\mbox{{\tiny spins}}}\,
\frac{1}{64}\sum_{\mbox{{\tiny colors}}}
|{\cal M}'(gg\to\chi_{cJ})_{q=0}|^2  \nonumber \\ && \times
 \F_g(x_1,k_{1t}^2,\mu^2)\,\F_g(x_2,k_{2t}^2,\mu^2)\,\, \label{ME_chi}
dk_{1t}^2\, dk_{2t}^2\, dy_{\chi}\,
\frac{d\phi_1}{2\pi}\,\frac{d\phi_2}{2\pi} .
\end{eqnarray}
%
%The explicit expressions for the parton level matrix elements
%$|{\M}(gg\to\chi_{cJ})|^2$ are presented in Ref. \cite{J_Tev}
%in the form of the expressions for the Feynman diagrams.
The squares of the matrix elements, as being too lengthy, are not
presented there but the full fortran code is available on request.
The numerical value of the wave function is taken from the potential 
model \cite{EicQui}: $|\R'_{\chi}(0)|^2 = 0.075$ GeV$^5$.
The decay branchings to \J meson are known to be \cite{PDG}
 $Br(\chi_{cJ}\to J/\psi\g)$ = 0.013, 0.35, and 0.20 
%\marginpar{$\surd$ 5}
%$Br(\chi_{cJ}\to J/\psi\g)$ = 0.006, 0.35, and 0.135 
for $J$ = 0, 1, and 2, respectively.
Here, the off-shell gluon flux factor is defined as 
%\marginpar{$\surd$ 6}
$F=2\lambda^{1/2}(\hat{s},k_1^2,k_2^2)$, according to the general
definition given by Eq.(2.3) in Ref. \cite{BycKaj}. 
For all other subprocesses one can use the approximations
$\lambda^{1/2}(\hat{s},k_1^2,k_2^2)\simeq\hat{s}\simeq x_1x_2s$, but they
are not suitable for the present case because the invariant mass of the 
final state is small and the difference between $\hat{s}\equiv m_\chi^2$ 
and $x_1x_2s\equiv m_{\chi,t}^2 = m_{\chi}^2+p_t^2$ can make pronounced
effect on the $p_T$ spectrum. The numerical accuracy of the above
definition was tested in a 
toy calculation regarding the leptonic production of \cj mesons via 
photon-photon fusion: $e+e \to e'+e'+\chi_c$. We have compared the exact 
$\O(\alpha^4)$ result with a number of calculations based on Equivalent 
Photon Approximation and using different definitions of the effective 
photon flux (such as $F=2\hat{s}$, $F=2x_1x_2s$, etc.). We find that the 
"$\lambda^{1/2}$" definition is in the best agreement with exact 
calculation.

For the production of beauty quarks in (\ref{bb}) we have
\begin{eqnarray}
&& d\sigma(pp\to b\bar{b}X)=
\frac{4\pi\alpha_s^2}{\hat{s}^2}\,
\frac{1}{4}\sum_{\mbox{{\tiny spins}}}\,
\frac{1}{64}\sum_{\mbox{{\tiny colors}}}
|{\cal M}(gg\to b\bar{b})|^2  \nonumber \\ && \times
 \F_g(x_1,k_{1t}^2,\mu^2)\,\F_g(x_2,k_{2t}^2,\mu^2)\,\, \label{ME_bb}
dk_{1t}^2\, dk_{2t}^2\, dp_{b T}^2\, dy_{b}\, dy_{\bar{b}}\,
\frac{d\phi_1}{2\pi}\,\frac{d\phi_2}{2\pi}\,\frac{d\phi_b}{2\pi}.
\end{eqnarray}
The explicit expressions for the parton level matrix elements
$|{\M}(gg\to b\bar{b})|^2$ can be found elsewhere \cite{bbar}.
In calculations the $b$-quark mass was set to $m_b$ = 4.5 GeV.
Further on, the produced $b$-quarks undergo fragmentation into
$B$-mesons according to the Peterson fragmentation function 
\cite{Peterson} with $\epsilon$=0.006. The outgoing $B$-mesons undergo then
a decay according to the three body decay mode $B\to J/\psi+K+\pi$,
to which the net effective branching fraction \cite{PDG} was attributed:
$Br(b\to J/\psi X)=1.16\%$ (resp., $Br(b\to\psi' X)=0.48\%$). 
%\marginpar{$\surd$ 7}
This decay mode was taken as a typical representative for all $B$-meson
decays. As the decay matrix elements are unknown, the decays were 
generated according to the phase space. However, the fine details of
fragmentation and decay are rather unimportant for our purposes, because
$b$-quarks play only marginal role at RHIC energies, except large
transverse momenta of $J/\psi$ or $\psi'$. We shall discuss the region
of the large transverse momenta somewhat later.

Finally, for the charm-associated production (\ref{jcc}) we write
\begin{eqnarray}
&& d\sigma(pp\to\psi c\bar{c}X)=
\frac{\alpha_s^4}{4\hat{s}^2}\, |\R(0)|^2
\frac{1}{4}\sum_{\mbox{{\tiny spins}}}\;
\frac{1}{64}\sum_{\mbox{{\tiny colors}}}
|{\cal M}(gg\to\psi c\bar{c})|^2 \nonumber \\ && \times 
\F_g(x_1,k_{1t}^2,\mu^2)\,\F_g(x_2,k_{2t}^2,\mu^2)\,
dk_{1t}^2 dk_{2t}^2 dp_{\psi T}^2 dp_{cT}^2 dy_\psi dy_c dy_{\bar{c}}
\frac{d\phi_1}{2\pi} \frac{d\phi_2}{2\pi}
\frac{d\phi_\psi}{2\pi} \frac{d\phi_c}{2\pi}. \label{ME_jcc}
\end{eqnarray}
The explicit expressions for the parton level matrix elements
$|{\M}(gg\to\psi c\bar{c})|^2$ as well as detailed description of the
kinematics are presented in Ref. \cite{JCC}.

To close the description of the production mechanisms, we wish to state 
that we do not consider explicitly color-octet contributions
in the present analysis. 
%as we believe that this model has already demonstrated its incapacity.
In fact, we know no data which would clearly manifest the
presence of color-octet contributions. On the contrary, the numerical
fits of the color-octet matrix elements based on the Tevatron and HERA
data are incompatible with each other. Moreover, a conflict
between the model predictions and the data on \J spin alignment indicate
that the production of vector quarkonia is certainly not dominated by the
color-octet mechanism. Some small contribution is not excluded
but cannot be calculated from first principles.

%------------------------------------------------------
\subsection{Unintegrated gluon distributions}
%------------------------------------------------------

In general, there are no simple relations between unintegrated
and integrated parton distributions.
Some of UGDFs in the literature are obtained based on familiar
collinear distributions, some are obtained by solving evolution
equations, some are just modelled or some are even parametrized.
A brief review of unintegrated gluon distributions (UGDFs) that will
be used also here can be found in Ref.\cite{LS06}.

At very low $x$ the unintegrated gluon distributions are believed
to fulfil BFKL equation \cite{BFKL}.
Here in the practical applications we shall use a simple
parametrization \cite{ELR96} for the numerical solution \cite{AKMS94}
and use acronym BFKL.
Another distribution closely related to the BFKL approach was
constructed by Bl\"umlein \cite{Bluem}. 
%In this approach $\alpha_s$ is a free parameter.

At large energies (small $x$) one expects in addition
saturation effects due to gluon recombinations.
A simple parametrization of unintegrated gluon distribution in the proton
can be obtained based on the Golec-Biernat--W\"usthoff
parametrization of the dipole-nucleon cross section with
parameters fitted to the HERA data.
The dipole-nucleon cross section can be transformed to corresponding
unintegrated gluon distribution.
The resulting gluon distribution can be found in \cite{GBW_glue}.
In the following we call it GBW UGDF for brevity.
Another parametrization, also based on the idea of gluon
saturation, was proposed in \cite{KL01}.
In contrast to GBW approach \cite{GBW_glue}, where
the dipole-nucleon cross section is parametrized,
in the Kharzeev-Levin (KL) approach it is the unintegrated gluon 
distribution which is parametrized. More details can be found in 
Ref.\cite{LS06}. 

Another useful parametrization, which describes the HERA
data, and therefore is valid for 10$^{-4}$ $< x <$ 10$^{-2}$ was
constructed by Ivanov and Nikolaev (IN) \cite{IN02}. We refer the reader
for details to the original paper. 

%We shall not repeat more details concerning those UGDFs here.

In some of approaches one imposes the following relation
between the standard collinear distributions and UGDFs:
\begin{equation}
g(x,\mu^2) = \int_0^{\mu^2} f_g(x,{\mathrm {\bf k}}_t^2,\mu^2)
\frac{d{\mathrm {\bf k}}_t^2}{{\mathrm {\bf k}}_t^2}   \; .
\end{equation}
%---------------

Due to its simplicity the Gaussian smearing of initial transverse momenta
is a good and popular reference for other approaches. It allows to study
phenomenologically the role of transverse momenta in several
high-energy processes.
We define a simple unintegrated gluon distribution:
\begin{equation}
{\cal F}_{g}^{Gauss}(x,k_t^2,\mu_F^2) = x g_{i}^{coll}(x,\mu_F^2)
\cdot f_{Gauss}(k_t^2) \; ,
\label{Gaussian_UPDFs}
\end{equation}
where $g^{coll}(x,\mu_F^2)$ is a standard collinear (integrated)
gluon distribution and $f_{Gauss}(k_t^2)$
is a Gaussian two-dimensional function:
\begin{equation}
f_{Gauss}(k_t^2) = \frac{1}{2 \pi \sigma_0^2}
\exp \left( -k_t^2 / 2 \sigma_0^2 \right) / \pi \; .
\label{Gaussian}
\end{equation}
The UGDF defined by Eq.(\ref{Gaussian_UPDFs}) and (\ref{Gaussian})
is normalized such that:
\begin{equation}
\int {\cal F}_{g}^{Gauss}(x,k_t^2,\mu_F^2) \; d k_t^2 = 
x g_{i}^{coll}(x,\mu_F^2) \; .
\label{Gaussian_normalization}
\end{equation}

At small values of $x$ the unintegrated gluon distribution can be 
obtained from integrated distribution as \cite{GLR83}:
%\marginpar{$\surd$ 8}
%
\begin{equation}
{\cal F}(x,k_t^2) = \frac{d (x g(x,\mu^2))}{d \mu^2} |_{\mu^2=k_t^2} \; .
\end{equation}
This method cannot be directly used at small transverse momenta (small
factorization scales) and must be supplemented by a further prescription.
One possible prescription is
freezing of the gluon distribution at $k_t^2 < \mu_{fr}^2$, another is
a shift of the scale: $\mu^2 \to \mu^2 + \mu_s^2$.
Of course $\mu_{fr}^2$ and $\mu_s^2$ are bigger than the lowest possible
scale for standard collinear distributions.
This method cannot be also applied at larger $x$ as here the scalling
violation reverses and negative values are obtained.

At intermediate and large $x$ more careful methods must be used.
Kwieci\'nski has shown that the evolution equations
for unintegrated parton distributions take a particularly
simple form in the variable conjugated to the parton transverse momentum.
In the impact-parameter space, the Kwieci\'nski equation
takes the following simple form \cite{kwiecinski}
\begin{equation}
\begin{split}
\frac{\partial{\tilde f_{NS}(x,b,\mu^2)}}{\partial \mu^2} &=
\frac{\alpha_s(\mu^2)}{2\pi \mu^2}  \int_0^1dz  \, P_{qq}(z)
\bigg[\Theta(z-x)\,J_0((1-z) \mu b)\,
{\tilde f_{NS}\left(\frac{x}{z},b,\mu^2 \right)}
\\&- {\tilde f_{NS}(x,b,\mu^2)} \bigg]  \; , \\
\frac{\partial{\tilde f_{S}(x,b,\mu^2)}}{\partial \mu^2} &=
\frac{\alpha_s(\mu^2)}{2\pi\mu^2} \int_0^1 dz
\bigg\{\Theta(z-x)\,J_0((1-z) \mu b)\bigg[P_{qq}(z)\,
 {\tilde f_{S}\left(\frac{x}{z},b,\mu^2 \right)}
\\&+ P_{qg}(z)\, {\tilde f_{G}\left(\frac{x}{z},b,\mu^2 \right)}\bigg]
 - [zP_{qq}(z)+zP_{gq}(z)]\,
{\tilde f_{S}(x,b,\mu^2)}\bigg\}  \; ,
 \\
\frac{\partial {\tilde f_{G}(x,b,\mu^2)}}{\partial \mu^2}&=
\frac{\alpha_s(\mu^2)}{2\pi \mu^2} \int_0^1 dz
\bigg\{\Theta(z-x)\,J_0((1-z) \mu b)\bigg[P_{gq}(z)\,
{\tilde f_{S}\left(\frac{x}{z},b,\mu^2 \right)}
\\&+ P_{gg}(z)\, {\tilde f_{G}\left(\frac{x}{z},b,\mu^2 \right)}\bigg]
-[zP_{gg}(z)+zP_{qg}(z)]\, {\tilde f_{G}(x,b,\mu^2)}\bigg\} \; .
\end{split}
\label{kwiecinski_equations}
\end{equation}
We have introduced here the short-hand notation
\begin{equation}
\begin{split}
\tilde f_{NS}&= \tilde f_u - \tilde f_{\bar u}, \;\;
                 \tilde f_d - \tilde f_{\bar d} \; ,  \\
\tilde f_{S}&= \tilde f_u + \tilde f_{\bar u} + 
                \tilde f_d + \tilde f_{\bar d} + 
                \tilde f_s + \tilde f_{\bar s} \; . 
\end{split}
\label{singlet_nonsinglet}
\end{equation}
The unintegrated parton distributions in the impact factor
representation are related to the familiar collinear distributions
as follows
\begin{equation}
\tilde f_{k}(x,b=0,\mu^2)=\frac{x}{2} p_k(x,\mu^2) \; .
\label{uPDF_coll_1}
\end{equation}
On the other hand, the transverse momentum dependent UPDFs are related
to the integrated parton distributions as
\begin{equation}
x p_k(x,\mu^2) =
\int_0^{\infty} d k_t^2 \; f_k(x,k_t^2,\mu^2) \; .
\label{uPDF_coll_2}
\end{equation}

The two possible representations are interrelated via Fourier-Bessel
transform
\begin{equation}
  \begin{split}
    &f_k(x,k_t^2,\mu^2) =
    \int_{0}^{\infty} db \;  b J_0(k_t b)
    {{\tilde f}_k(x,b,\mu^2)} \; ,
    \\
    &{\tilde f}_k(x,b,\mu^2) =
    \int_{0}^{\infty} d k_t \;  k_t J_0(k_t b)
    {f_k(x,k_t^2,\mu^2)} \; .
  \end{split}
\label{Fourier}
\end{equation}
The index $k$ above numerates either gluons ($k$=0), quarks ($k >$ 0) or
antiquarks ($k <$ 0).

The perturbative solutions ${\tilde f}_k^{pert}(x,b,\mu_F^2)$ do not
include nonperturbative effects such as, for instance,
intrinsic transverse momenta of partons in colliding hadrons.
One of the reasons is e.g. internal motion of constituents of the proton. 
In order to include such effects we modify the perturbative
solution $\tilde{f}_g^{pert}(x,b,\mu_F^2)$
and write the modified parton distributions
$\tilde{f}_g(x,b,\mu_F^2)$ in the simple factorized form
\begin{equation}
\tilde{f}_g(x,b,\mu_F^2) = \tilde{f}_g^{pert}(x,b,\mu_F^2)
 \cdot F_g^{np}(b) \; .
\label{modified_uPDFs}
\end{equation}
In the present study we shall use the following functional
form for the nonperturbative form factor
\begin{equation}
F_k^{np}(b) = F^{np}(b) = \exp\left(- \frac{b^2}{4 b_0^2}\right) \; .
%\text{or} \; \exp \left( - \frac{b}{b_e} \right)
\label{formfactor}
\end{equation}
In Eq.(\ref{formfactor}) $b_0$ is the only free parameter.

While physically $f_k(x,k_t^2,\mu^2)$ should be positive,
there is no obvious reason for such a limitation for
$\tilde f_k(x,b,\mu^2)$.

In the following we use leading-order parton distributions
from Ref.\cite{GRV98} as the initial condition for QCD evolution.
The set of integro-differential equations in $b$-space
was solved by the method based on the discretisation made with
the help of the Chebyshev polynomials (see \cite{kwiecinski}).
Then the unintegrated parton distributions were put on a grid
in $x$, $b$ and $\mu^2$ and the grid was used in practical
applications for Chebyshev interpolation. 

For the calculation of the direct $J/\psi$ production here 
the parton distributions in momentum space are more useful.
This calculation requires a time-consuming multi-dimensional
integration. An explicit calculation of the Kwieci\'nski UPDFs
via Fourier transform for needed in the main calculation values of
$(x_1,k_{1,t}^2)$ and $(x_2,k_{2,t}^2)$ (see next section)
is not possible.
Therefore auxiliary grids of
the momentum-representation UPDFs are prepared before
the actual calculation of the cross sections.
These grids are then used via a two-dimensional interpolation
in the spaces $(x_1,k_{1,t}^2)$ and $(x_2,k_{2,t}^2)$
associated with each of the two incoming partons.

%The evolution of the parton cascade leads to a spread of
%the transverse momentum of the parton at the end of the cascade,
%the parton participating in a hard process.
%Let us define the following measure of the spread:
%
%\begin{equation}
%< k_t^2 >_{f_k}(x,\mu^2) \equiv 
%\int d k_t^2 \; k_t^2 \; f_k(x,k_t^2,\mu^2) /
%\int d k_t^2 \; f_k(x,k_t^2,\mu^2) \; .
%\end{equation}
%
%As an example in Fig.\ref{fig:ave_kt2_x} we show
%the spread, obtained for a few unintegrated parton distributions
%obtained by the Fourier transform of the impact factor distributions
%obtained themselves by solving the Kwieci\'nski equations,
%as a function of parton longitudinal momentum fraction.
%In this calculation the factorization scale was fixed at
%$\mu^2 =$ 100 GeV$^2$. 
%The Kwieci\'nski evolution leads to the increasing spread
%with decreasing longitudinal momentum fraction.
%The spread for different distributions is different.
%In the region of small $x$ the spread in $k_t^2$ for gluons is bigger
%than a similar spread for sea and valence quarks.
%This is very different than a corresponding spread for Gaussian
%distributions which is usually taken to be independent of $x$
%and parton species. In addition, the spread of $k_t^2$ for small
%values of $x$ is considerably larger than the nonperturbative
%spread of the initial Gaussian distributions, taken here identical for
%all species and encoded in the model parameter $b_0$.

The Kwieci\'nski unintegrated parton distributions were used recently in
applications to $c \bar c$ photoproduction \cite{LS04},
$c \bar c$ correlations in nucleon-nucleon collisions \cite{LS06},
production of gauge bosons \cite{KS04}, production of Standard Model
Higgs boson \cite{Higgs_LS06}, inclusive production of pions
\cite{CS05}, production of direct photons \cite{PS07}.
Good agreement with experimental data was obtained in the case
when the data exist.

In the approach of Ref. \cite{Bluem}, based on leading-order
perturbative solution of BFKL equation, the unintegrated gluon density
$\F_g(x,k_{t}^2,\mu^2)$  is calculated as a convolution of the ordinary
(collinear) gluon density $g(x,\mu^2)$ with universal weight factors:
\begin{equation} \label{conv}
 {\cal F}_g(x,k_{t}^2,\mu^2) = \int_x^1
 {\cal G}(\eta,k_{t}^2,\mu^2)\,
 \frac{x}{\eta}\,g(\frac{x}{\eta},\mu^2)\,d\eta,
\end{equation}
\begin{equation} \label{J0}
 {\cal G}(\eta,k_{t}^2,\mu^2)=\frac{\bar{\alpha}_s}{\eta\,k_{t}^2}\,
 J_0(2\sqrt{\bar{\alpha}_s\ln(1/\eta)\ln(\mu^2/k_{t}^2)}),
 \qquad k_{t}^2<\mu^2,
\end{equation}
\begin{equation}\label{I0}
 {\cal G}(\eta,k_{t}^2,\mu^2)=\frac{\bar{\alpha}_s}{\eta\,k_{t}^2}\,
 I_0(2\sqrt{\bar{\alpha}_s\ln(1/\eta)\ln(k_{t}^2/\mu^2)}),
 \qquad k_{t}^2>\mu^2,
\end{equation}
where $J_0$ and $I_0$ stand for Bessel functions (of real and imaginary
arguments, respectively), and $\bar{\alpha}_s=\alpha_s/3\pi$. The LO GRV
set \cite{GRV95} was used in our calculations as the input collinear 
density. Here the value of $\alpha_s$ and the scale $\mu^2$ are
parameters of the model. The resulting unintegrated gluon distributions
depend on them rather strongly.
Sometimes for brevity we shall denote the distribution from
Ref.\cite{Bluem} by JB.

%---------------------------------
\section{Results}
%---------------------------------

\begin{figure}[htb]    % 
 {\includegraphics[width=0.6\textwidth]{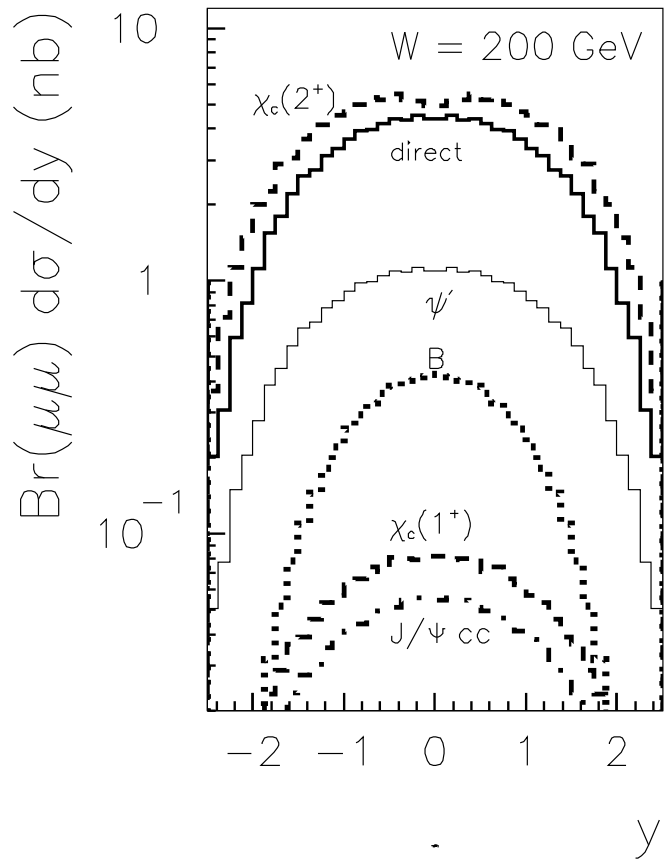}}
   \caption{\label{fig:all_y}
\small
Contributions of different mechanisms for the production of $J/\psi$
in $d \sigma / dy$ distributions. In this calculation we have used
simple ``derivative UGDF''.
}
\end{figure}

Now we shall compare contributions of different processes discussed
in the previous section. Here a Monte Carlo method based on the VEGAS
routine \cite{VEGAS} is used to allow an easy comparison of processes
with different number of particles in the final state.
In Fig. \ref{fig:all_y} we show contribution of different mechanisms
discussed above to the rapidity distributions of $J/\psi$ meson
for the RHIC energy. This calculation is based on so-called derivative
UGDFs, i.e. the ones obtained by differentiating the standard collinear
distributions (see the previous section). 

%The factorization scale was fixed as
%$\mu^2 = m_{t,J/\psi}^2$ for the direct color-singlet production,
%$\mu^2 = m_{t,\chi_c}^2$ for the reactions proceeding via radiative
%decays of $\chi_c$ mesons, $\mu^2 = ........$ for the open-charm
%associated reaction. In the case of weak decays of $B$ mesons
%.................................................................

\begin{figure}[htb]    % 
 {\includegraphics[width=0.6\textwidth]{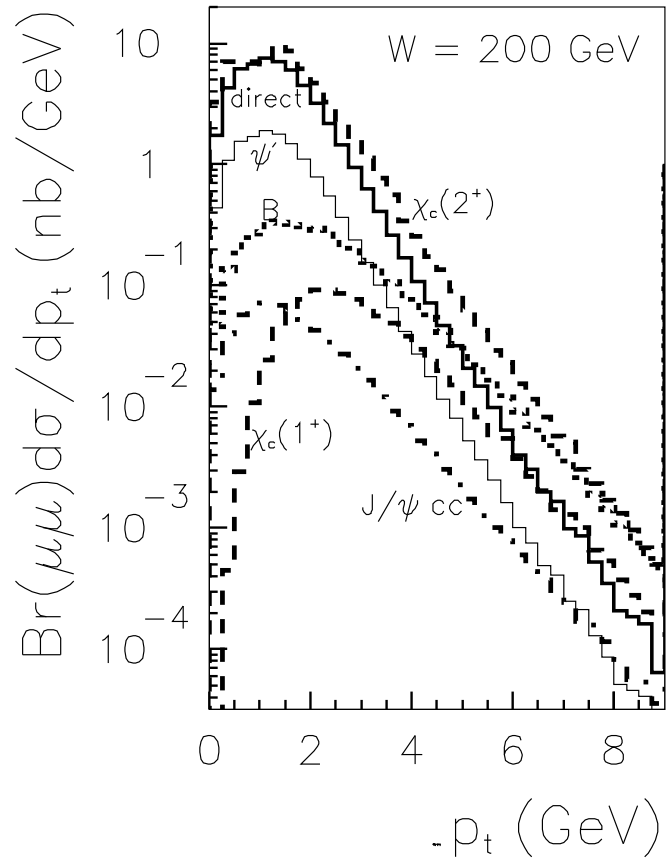}}
   \caption{\label{fig:all_pt}
\small
Contributions of different mechanisms for the production of $J/\psi$
in $d \sigma / dp_t$ distributions. In this calculation we have used
simple ``derivative UGDF''. The cross section is integrated over
the full range of rapidity.
}
\end{figure}

In Fig.\ref{fig:all_pt} we show corresponding contributions to the
transverse momentum distribution of the $J/\psi$ meson.
In this exploratory calculation the cross section is integrated over
the full range of rapidity. 
We obtain a rather surprising result that the sequential production of
$J/\psi$ mesons via radiative decays of $\chi_c$ mesons is comparable to
or even dominates over the direct color-singlet contribution almost in
the whole phase space.                               
%\marginpar{$\surd$ 10}
The reason can be seen in the fact that the production of \cj states
refers to much lower values of the final state invariant mass,
$m_\chi^2<<(p_\psi+p_ g)^2$, giving emphasis to the small $x$ region, 
where the gluon distributions are growing up. This property becomes even 
more pronounced as the 'direct' matrix element (\ref{direct}) vanishes 
when the emitted final gluon is soft.
Our conclusion on the relative size of the direct and indirect 
contributions is compatible with the preliminary estimates obtained by 
the CDF collaboration \cite{CDF3}.

%It is worth noting that the production of $\chi_c$ mesons can hardly
%be described in a consistent way within the collinear factorization
%scheme.                                              
%\marginpar{$\surd$ 11}
We wish to note now some difficulties of the standard collinear approach
for the $\chi_c$ mesons.
The leading order contribution coming from the subprocess (\ref{chi})
shows unphysical $\delta$-like $p_T$ spectrum. The usual excuse
that the particles produced at zero $p_T$ disappear in the beam pipe
and remain invisible does not work, because the decay products do have
nonzero $p_T$ and can be detected. At the same time,
introducing the next-to-leading contributions (i.e., the processes
with extra gluons in the final state) causes a problem of infrared
divergences, which need artificial tricks to regularize them.

It is well known that a large fraction of the $\psi'$ mesons
decays into channels with $J/\psi$ (BR = 0.56 \cite{PDG}).
This contribution was not considered in the literature
and requires a separate discussion.
The inclusive cross section for $\psi'$ can be calculated in exactly
the same way (color-singlet model) as the cross section for direct
$J/\psi$ meson production.
The decays of $\psi' \to J/\psi + X$ change the kinematics only slightly.
%......(wave function + branching ratio + decay kinematics)....................%......................\\
Finally the $\psi'$ contribution constitutes about 25 \%
of the direct (color-singlet) production.

Also the B-meson decay mechanism gives a sizeable contribution at large
transverse momenta.

Summarizing, at the RHIC energy the dominant production mechanisms
are radiative decays of $\chi_c(2^+)$ and direct color-singlet
mechanism. In the following we shall concentrate exclusively on these two 
dominant mechanisms.

\begin{figure}[htb]    % 
 {\includegraphics[width=0.6\textwidth]{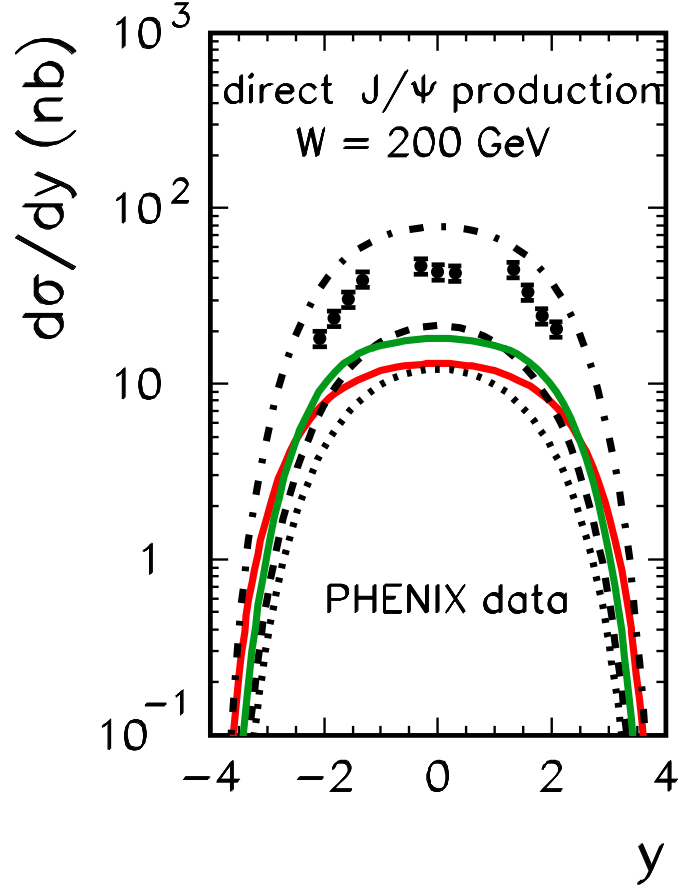}}
   \caption{\label{fig:dsig_dy_direct}
\small Direct color-singlet contribution to rapidity distribution of $J/\psi$
for different models of UGDFs.
The solid (red on line) curve corresponds to the Kwieci\'nski UGDF, the
dashed line to the Kharzeev-Levin UGDF, the dotted line to the BFKL
UGDF, the dash-dotted line to the Ivanov-Nikolaev UGDF and the grey solid
(green on line) curve to the Bluemlein UGDF.
The $\psi'$ contribution is not included here.
The new PHENIX data are shown as full circles.}
\end{figure}

\begin{figure}[htb]
 {\includegraphics[width=0.4\textwidth]{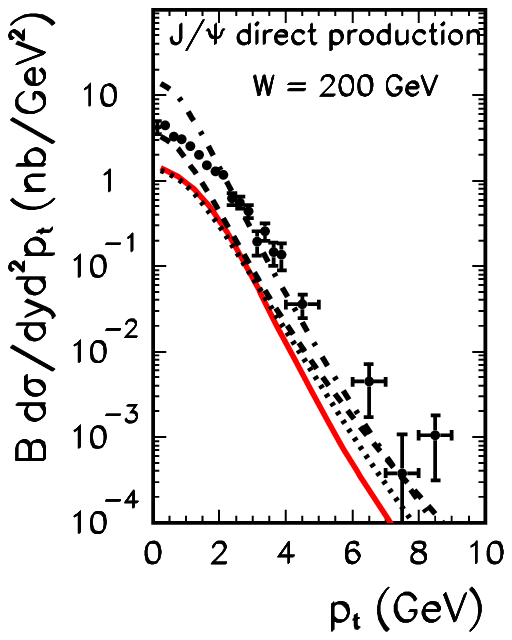}}
 {\includegraphics[width=0.4\textwidth]{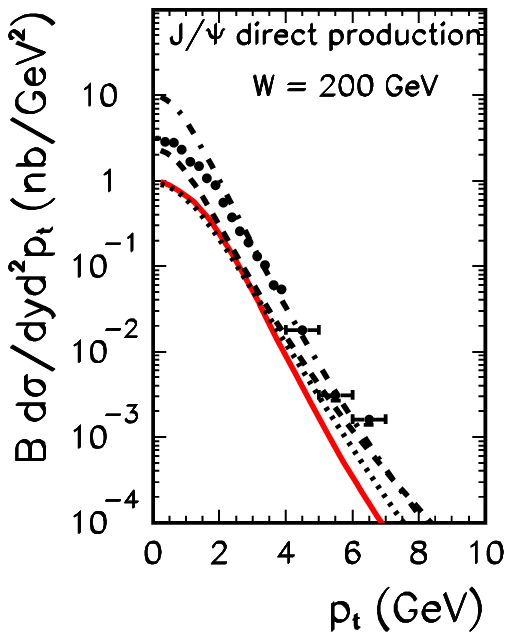}}
   \caption{\label{fig:dsig_dpt_direct}
\small Direct color-singlet contribution to transverse momentum distribution
of $J/\psi$ for different models of UGDFs for different intervals
in rapidity: (a) -0.35 $ < y <$ 0.35 (left panel), (b) 1.2 $ < |y| < $
2.2 (right panel). The meaning of the curves is the same as in Fig.
\ref{fig:dsig_dy_direct}. The $\psi'$ contribution is not included here.
The new PHENIX data are shown as full circles.}
\end{figure}

Let us start with color-singlet mechanism. In Fig.\ref{fig:dsig_dy_direct}
we present distributions in rapidity of $J/\psi$ produced by direct
color-singlet mechanism for different UGDFs. The distribution obtained
with Ivanov-Nikolaev (IN) glue exceeds the experimental PHENIX data
\cite{RHIC_JPSI}, while the other theoretical distributions are smaller
than experimental data. This is rather natural as contributions
of other mechanisms are not included.
The corresponding distributions in transverse momentum are shown in
Fig.\ref{fig:dsig_dpt_direct} for two different intervals in rapidity. 
Very similar distributions are obtained for mid- and intermediate rapidity
intervals. The result with Ivanov-Nikolaev UGDF exceeds the
experimental data in the region of small transverse momenta.
This is probably due to an extra nonperturbative contribution
at small gluon transverse momenta \cite{IN02}.

\begin{figure}[htb]    % 
 {\includegraphics[width=0.6\textwidth]{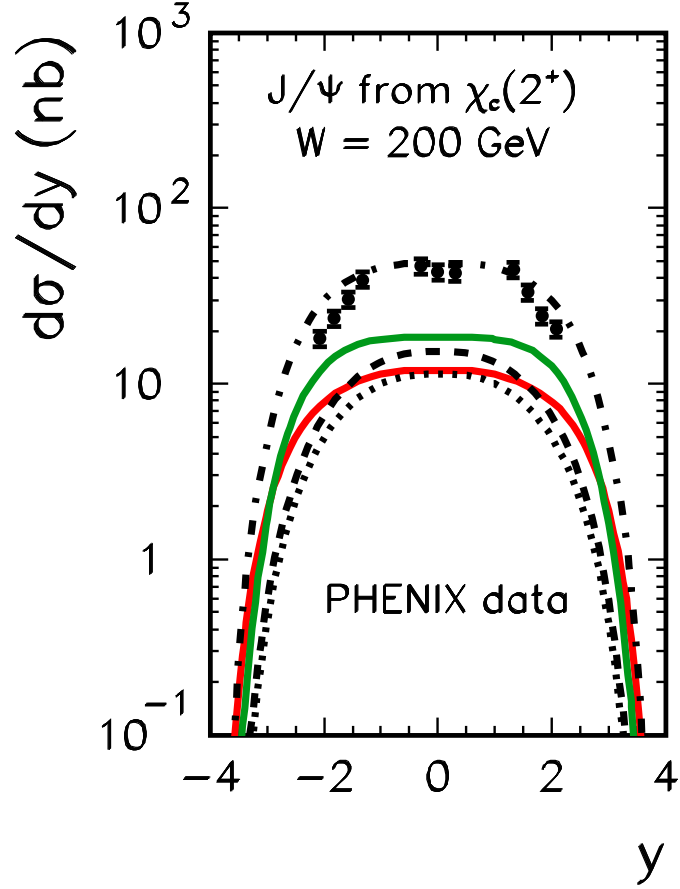}}
   \caption{\label{fig:dsig_dy_chic2}
\small $\chi_c$-decay contribution to rapidity distribution of $J/\psi$
for different models of UGDFs.
The meaning of the curves is the same as in Fig.
\ref{fig:dsig_dy_direct}.
 The new PHENIX data are shown as full
circles.}
\end{figure}

\begin{figure}[htb]    % 
 {\includegraphics[width=0.4\textwidth]{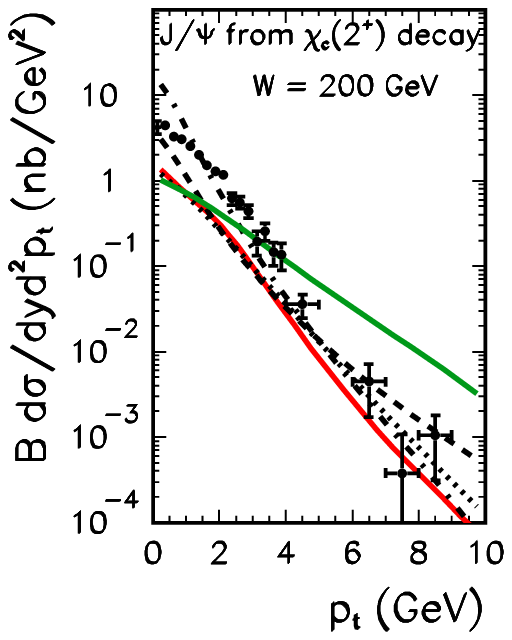}}
 {\includegraphics[width=0.4\textwidth]{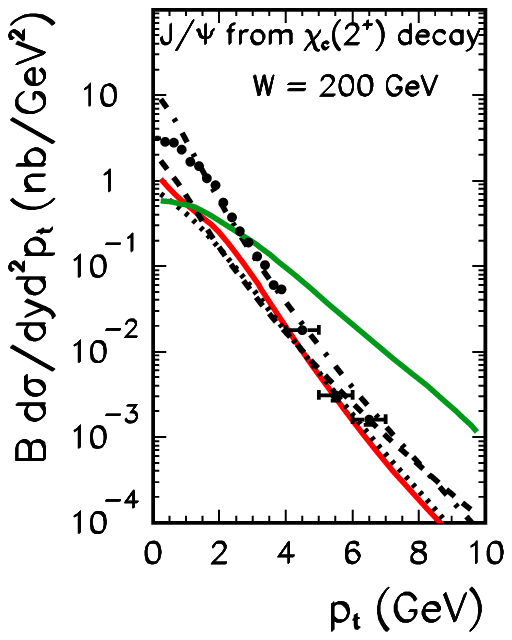}}
   \caption{\label{fig:dsig_dpt_chic2}
\small $\chi_c$-decay contribution to transverse momentum distribution
of $J/\psi$ for different models of UGDFs for different intervals
in rapidity: (a) -0.35 $ < y <$ 0.35 (left panel), (b) 1.2 $ < |y| < $
2.2 (right panel).
The meaning of the curves is the same as in Fig.
\ref{fig:dsig_dy_direct}.
The new PHENIX data are shown as full circles.}
\end{figure}

Now we shall show results obtained with different UGDFs for radiative
decays of $\chi_c(2^+)$. The rapidity distribution of corresponding
$J/\psi$ is shown in Fig.\ref{fig:dsig_dy_chic2}. Different UGDFs give a
similar result. The distributions obtained with Ivanov-Nikolaev UGDF
is slightly higher than those obtained with other distributions.
In Fig.\ref{fig:dsig_dpt_chic2} we show distributions in transverse
momentum of radiatively produced $J/\psi$. The differences in the
results for different UGDFs are up to a factor 2 or even larger. 
Again Ivanov-Nikolaev UGDF gives the highest cross section for small
transverse momenta. The Bluemlein UGDF shown intentionally
for large value of $\alpha_s = 0.6$ (solid grey, green on line) gives
completely wrong shape. The shape in this case depends strongly
on the value of $\alpha_s$. It would be much better for smaller
values of $\alpha_s$.

\begin{figure}[htb]    % 
 {\includegraphics[width=0.6\textwidth]{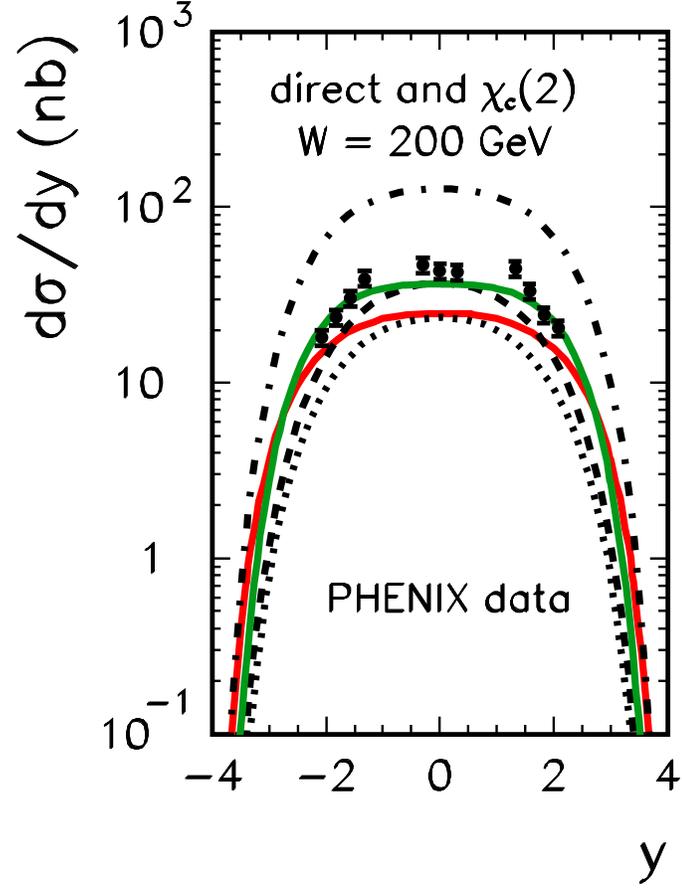}}
   \caption{\label{fig:dsig_dy_sum}
\small Direct color-singlet and $\chi_c(2^+)$ contributions to rapidity
distribution of $J/\psi$ for different models of UGDFs.
The meaning of the curves is the same as in Fig.
\ref{fig:dsig_dy_direct}.
The new PHENIX data are shown as full circles.}
\end{figure}

\begin{figure}[htb]    % 
 {\includegraphics[width=0.4\textwidth]{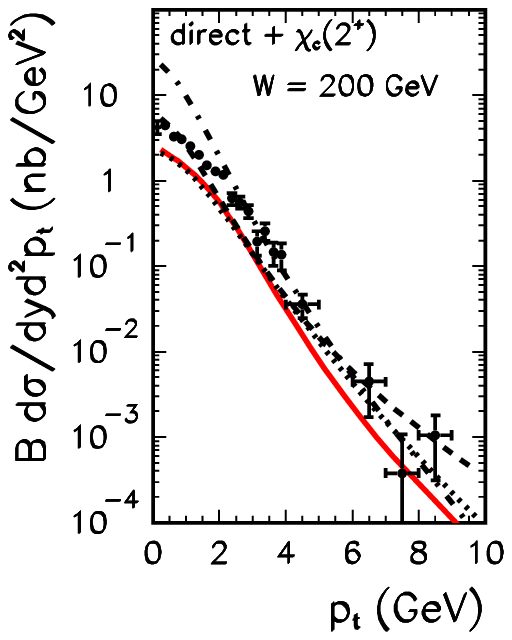}}
 {\includegraphics[width=0.4\textwidth]{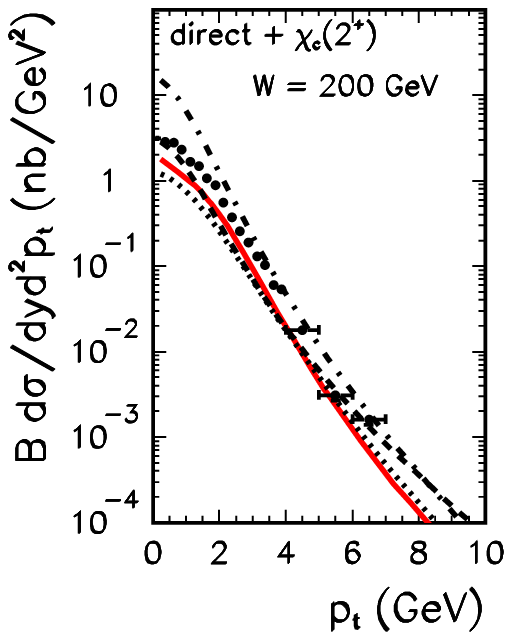}}
   \caption{\label{fig:dsig_dpt_sum}
\small Direct and $\chi_c$-decay contributions to transverse momentum
distribution of $J/\psi$ for different models of UGDFs for different intervals
in rapidity: (a) -0.35 $ < y <$ 0.35 (left panel), (b) 1.2 $ < |y| < $
2.2 (right panel). The meaning of the curves is the same as in Fig.
\ref{fig:dsig_dy_direct}.
The new PHENIX data are shown as full circles.}
\end{figure}

Finally we would like to show how the sum of the two dominant
contributions (direct color-singlet and radiative $\chi_c(2^+)$ decay)
compares with the experimental data from RHIC.
The distribution in rapidity is shown in Fig.\ref{fig:dsig_dy_sum}
and distributions in transverse momentum in Fig.\ref{fig:dsig_dpt_sum}.
The theoretical cross sections obtained with the Kwieci\'nski,
BFKL and Kharzeev-Levin UGDFs stay slightly below the experimental data.
This seems to be consistent with the fact that the smaller contributions
discussed in Fig.\ref{fig:all_y} and Fig.\ref{fig:all_pt} are not
included here. They are expected to produce contributions of the order
of 20--30 \% (see Figs.\ref{fig:all_y},\ref{fig:all_pt}).

\begin{figure}[htb]   
 {\includegraphics[width=0.4\textwidth]{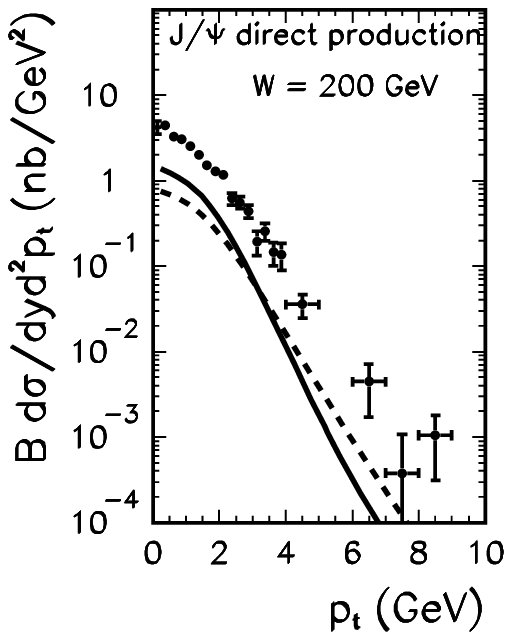}}
 {\includegraphics[width=0.4\textwidth]{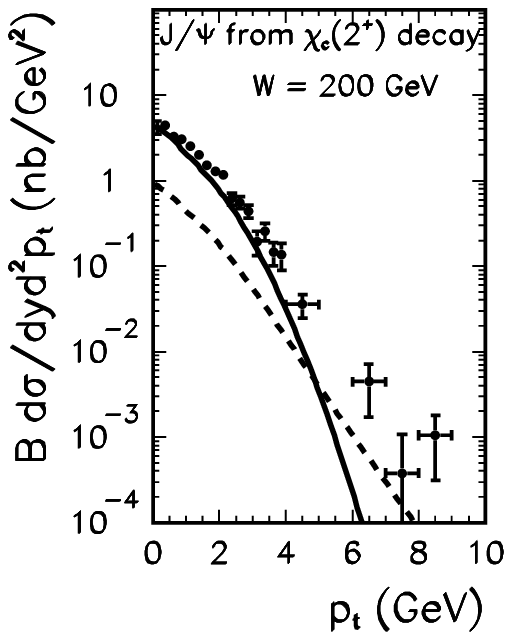}}
   \caption{\label{fig:inv_pt_kwiec_scales}
\small Factorization scale dependence of the transverse momentum distribution
for Kwieci\'nski UGDF. The mid rapidity range
-0.35 $ < y <$ 0.35 was taken as an example.
The left panel is for direct production and the right panel for the
$\chi_c(2^+)$ decay mechanism.
The solid and dashed curves are for $\mu^2$ = 10 GeV$^2$
and for $\mu^2$ = 100 GeV$^2$, respectively. In this calculation
$b_0$ = 1 GeV$^{-1}$.
}
\end{figure}

At the RHIC energy $W$ = 200 GeV the longitudinal momentum fractions
of the order $x \sim$ 10$^{-2}$ -- 10$^{-1}$ come into game.
This is the place where application of many UGDFs may be questionable.
Let us concentrate now on Kwieci\'nski parton distributions, which are
constructed for the region of $x$ under discussion.
In the left panel of Fig.\ref{fig:inv_pt_kwiec_scales} we show invariant
cross section for the direct component as a function of $J/\psi$
transverse momentum $p_t$ for mid rapidity range
-0.35 $< y <$ 0.35.
We show results for different factorization scales: $\mu^2$ = 10 GeV$^2$
(solid) and $\mu^2$ = 100 GeV$^2$ (dashed).
% as well as for running scale
%$\mu^2 = (m_{1t}^2+m_{2t}^2)/2$ (solid).
In the right panel of Fig.\ref{fig:inv_pt_kwiec_scales} we show similar
result for $J/\psi$ coming from the decays of the $\chi_c(2^+)$.
Here the result depends more strongly on the choice of the scale.
The solid line here corresponds to running factorization scale:
$\mu^2 = m_t^2 = m_{\chi_c(2^+)}^2 + p_t^2$.
 
\begin{figure}[htb]   
 {\includegraphics[width=0.4\textwidth]{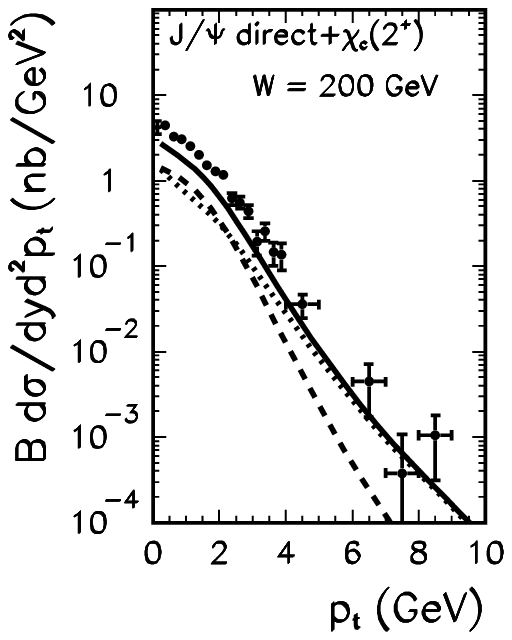}}
 {\includegraphics[width=0.4\textwidth]{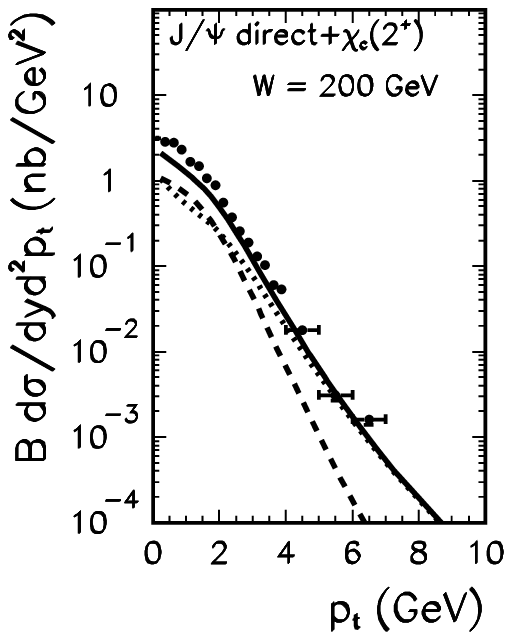}}
   \caption{\label{fig:inv_pt_kwiec_running_scale}
\small Invariant cross section for the Kwieci\'nski UGDF with running
scale.
The left panel is for central rapidity range (-0.35 $< y <$ 0.35) and
the right panel for intermediate rapidity range (1.2 $< y <$ 2.2).
The direct contribution is denoted by the dashed line,
the $\chi_c(2^+)$-decay contribution by the dotted line
and the sum of both by the solid line.
}
\end{figure}

In Fig.\ref{fig:inv_pt_kwiec_running_scale} we compare the sum of
both processes calculated with running factorization scale with the
PHENIX experimental data. 
%A rather good description of the data is obtained. 
The calculation underestimates the data at small transverse
momenta. This is most probably due to the omission of other components,
especially the $\psi'$-decay component.

\begin{figure}[!htb]
 {\includegraphics[width=0.4\textwidth]{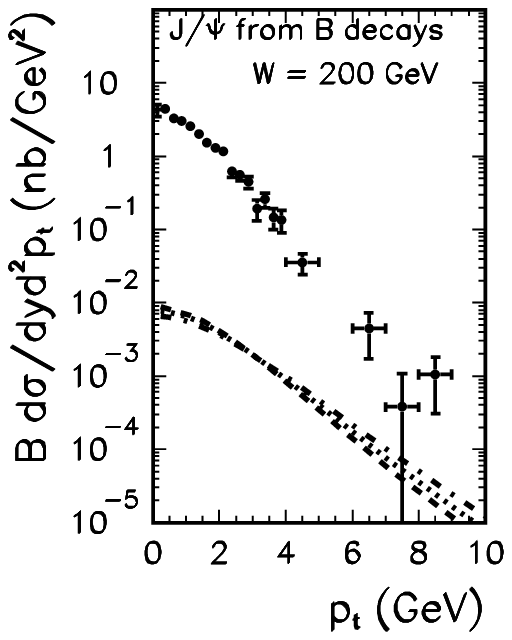}}
 {\includegraphics[width=0.4\textwidth]{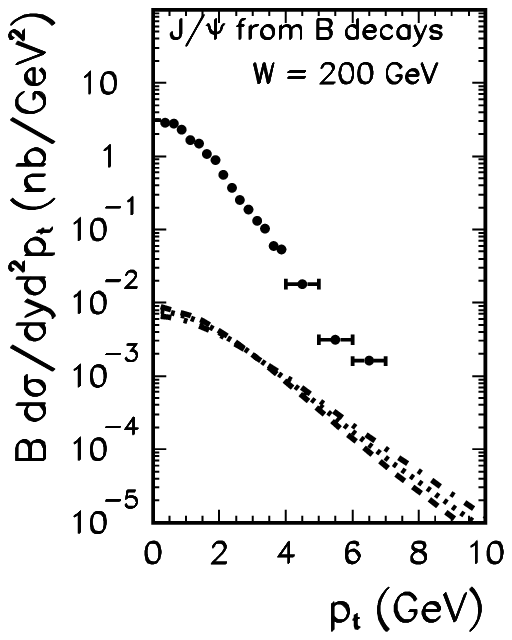}}
   \caption{\label{fig:inv_pt_bdecays}
\small Invariant cross section for $J/\psi$ from decays of the $B$
mesons as a function of $p_t$ for midrapidity and intermediate rapidity
intervals.
Kwiecinski UGDFs are used with factorization scale $\mu^2 = 4 m_b^2$.
Different decay models are described in the text.
}
\end{figure}

Let us concentrate now on the region of large transverse momenta of $J/\psi$.
In Fig.\ref{fig:inv_pt_bdecays} we show the contribution of $J/\psi$
from decays of the $B$ and $\bar B$ mesons.
The cross section for the $b \bar b$ is obtained with
the Kwieci\'nski UGDF (fixed factorization and renormalization scales,
$\mu^2 = 4 m_b^2$) within the $k_t$-factorization approach.
The details of the calculation can be found in Ref.\cite{LS06}.
In the present illustrative calculation we neglect hadronization, i.e.
we assume that the distribution of $B$ ($\bar B$) mesons is the same as the
distribution of $b$ ($\bar b$) quarks. This seems justified for heavy
quark to heavy meson transitions. There are several decay channels
with final state $J/\psi$.
The inclusive branching ratio is known experimentally BR = 1.09\% \cite{PDG}.
However, the momentum distribution of $J/\psi$ in the $B$ meson 
center-of-mass system was not yet measured 
\footnote{In principle it could be measured
by the Belle or Babar collaborations.}. Here, in order to demonstrate
the dependence on the details of the decay, we consider 3 academic models
of the decays: (a) uniform distribution in $p^*$ (momentum of $J/\psi$
in the meson rest frame) in the interval
(0,$p_{max}$) -- dashed line, (b) uniform distribution inside the sphere
with radius $p_{max}$ -- dotted line, (c) distribution on the sphere
with radius $p_{max}$ -- dash-dotted line.
Here $p_{max}$ is the momentum obtained assuming a two-body decay:
$B \to J/\psi X$. We assume the effective mass of the state $X$ to be
$m_X$ = 0.5 GeV.
As can be seen from the figure the $B$ decays become an important
ingredient at larger transverse momenta. There is relatively mild dependence
on the details of the decay. However, these details may become
important with a better statistics, when $J/\psi$ with $p_t >$ 10 GeV
will be measured. The present estimate of the B-decay contribution
may be an underestimation because of the two following reasons: 
(a) it is based on leading-order approach, 
(b) choice of the renormalization scale (see above)
 \footnote{Often in the $k_t$-factorization approach transverse momenta
squared of initial gluons are taken as the scale of running
$\alpha_s$.}.
Therefore at presently measured maximal transverse momenta
of $J/\psi$ $p_t \sim$ 8 GeV the B-decay contribution at the level of
20-30 \% is not excluded.

\begin{figure}[htb]   
 {\includegraphics[width=0.4\textwidth]{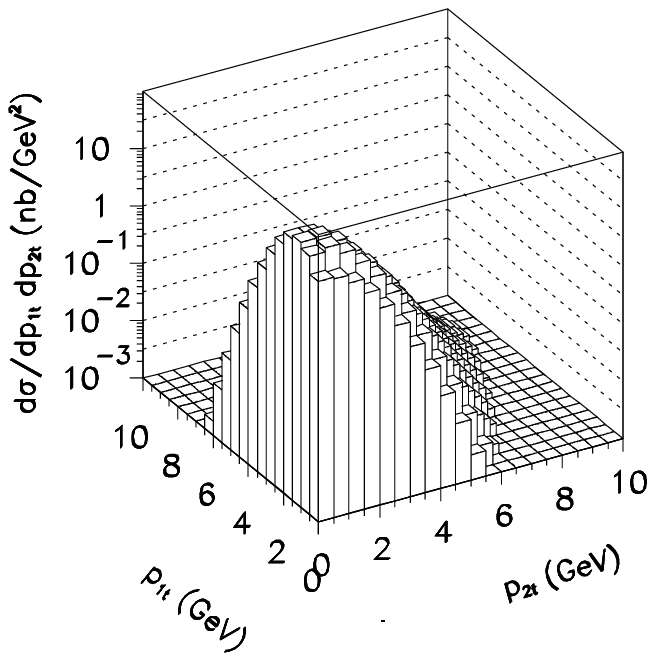}}
 {\includegraphics[width=0.4\textwidth]{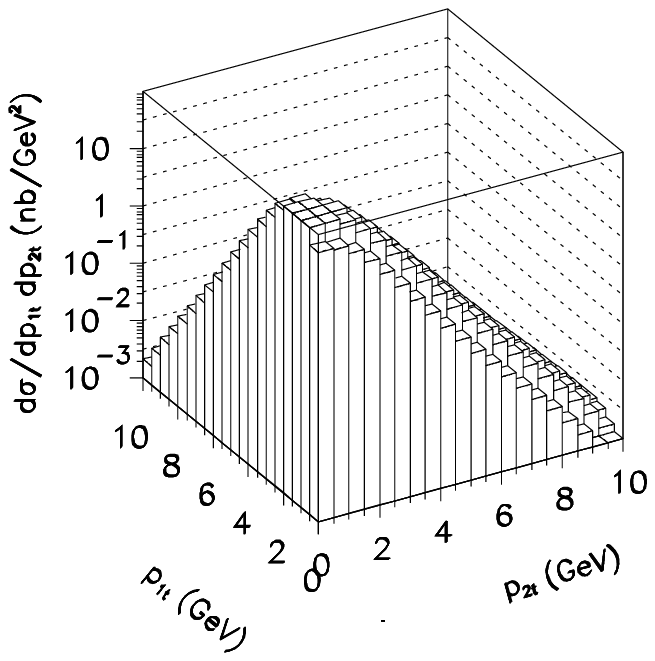}}
   \caption{\label{fig:map_p1tp2t_kwiec}
\small Factorization scale dependence of the $p_{J/\psi,t} \times p_{g,t}$
distribution for the Kwieci\'nski UGDF.
The left panel is for $\mu^2$ = 10 GeV$^2$ and the right panel is
for $\mu^2$ = 100 GeV$^2$.
}
\end{figure}

Let us concentrate now on correlations between produced $J/\psi$
and associated gluon(s).
In Fig.\ref{fig:map_p1tp2t_kwiec} we present two-dimensional
distribution in transverse momentum of $J/\psi$ ($p_{1t}$) and
transverse momentum of the associated (the gluon related with the matrix
element) gluon ($p_{2t}$) for two different scales of the Kwieci\'nski
UGDF. The bigger the scale is, the bigger is
the spread of the cross section in the ($p_{1t}, p_{2t}$) space.
This can be understood by the fact that the bigger scales means more
gluonic emissions which statistically means the bigger spread.
%                                                    \marginpar{$\surd$ 12}
%We wish to point out that the cross section at small transverse momenta
%diminishes and does not show any singularity which is present
%in the collinear approach. This is due to the behaviour of the matrix
%element for the corresponding $2 \to 2$ partonic subprocess.
This figure is rather of academic value as in practice there are
also gluons emitted in the process of the ladder-type emissions.
Strictly speaking, the latter have to be described using a full gluon 
evolution generator. On the other hand, the relevant effects can also be 
estimated in an approximate way, as follows.      
%\marginpar{$\surd$ 13}
On the average, the gluon transverse momentum increases from the proton
line towards the hard interaction block (although there is no strict 
ordering in the transverse momentum in BFKL equation). So, it is most
likely, that the last gluon in the parton ladder has the largest $k_t$ 
value. As a rough approximation, one can neglect the transverse momenta 
of all the other emitted gluons (note that the evolution is in the 
$\log(k_t)$ space rather than $k_t$ space) and use the conservation law 
in the last splitting vertex to set the $k_t'$ of the emitted gluon 
opposite to the $k_t$ of the gluon entering the partonic matrix element: 
$\vec{k}_t'\simeq -\vec{k}_t$. The latter is known from the unintegrated
gluon distribution. This trick gives an estimate for the transverse momentum 
of the final state gluon jet.

\begin{figure}[htb]   
 {\includegraphics[width=0.4\textwidth]{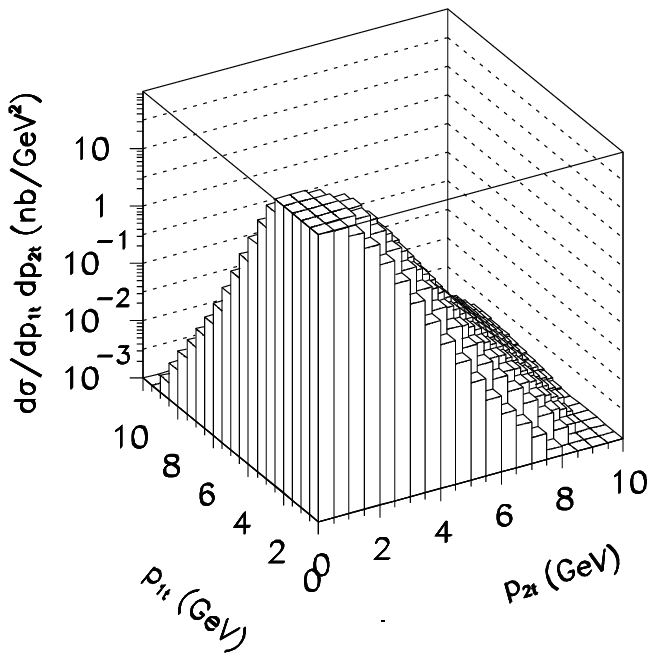}}
 {\includegraphics[width=0.4\textwidth]{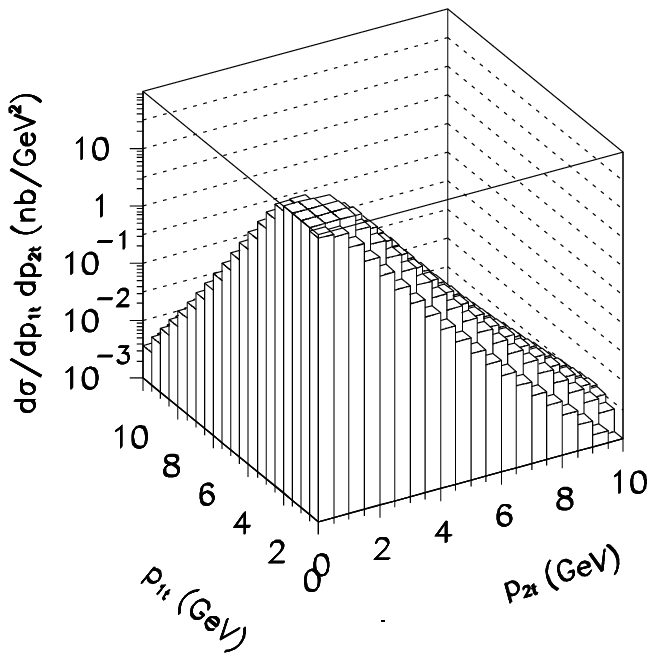}}
 {\includegraphics[width=0.4\textwidth]{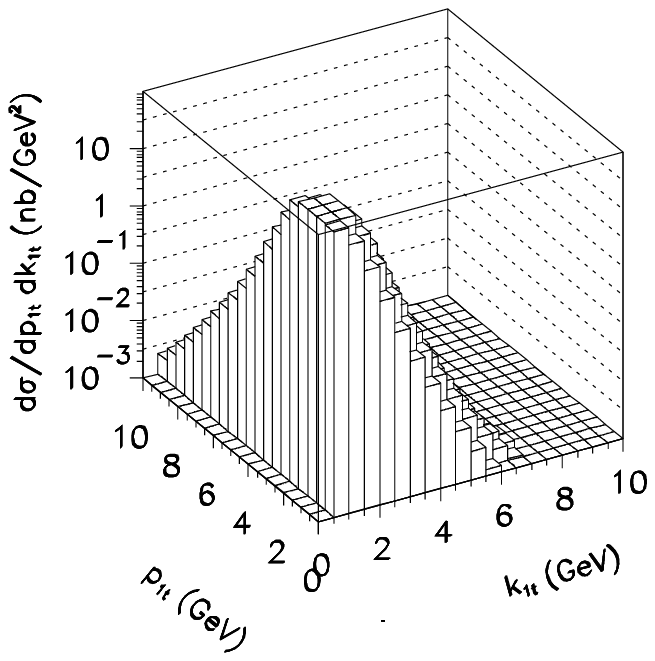}}
 {\includegraphics[width=0.4\textwidth]{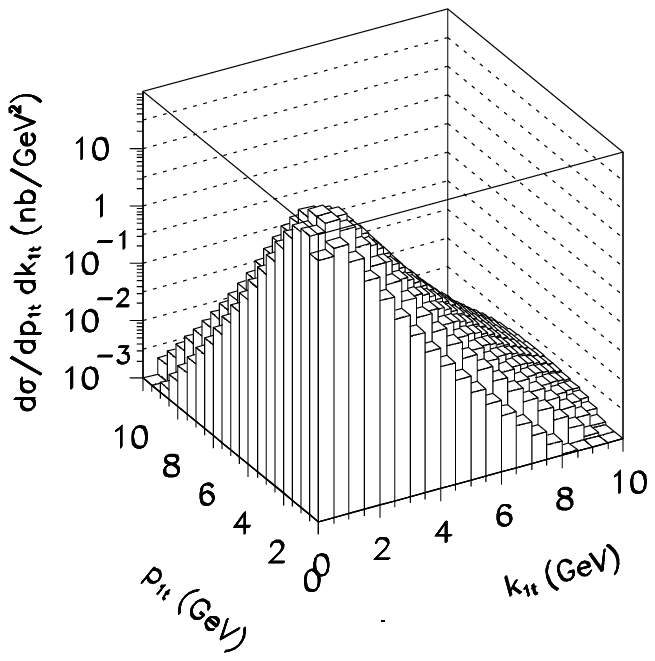}}
   \caption{\label{fig:map_jpsi_gluon}
\small
Two-dimensional distribution of the $J/\psi$ and gluon transverse
momenta. The {\it left-top} panel is for Kwieci\'nski UGDF (running scale)
and matrix element gluon, the {\it right-top} panel is for the BFKL UGDF
and matrix element gluon, the {\it left-bottom} panel is for the Kwieci\'nski
UGDF (running scale) and "last from the ladder" gluon and the
{\it right-bottom} panel is for the BFKL UGDF and "last from the ladder"
gluon.
}
\end{figure}

In Fig.\ref{fig:map_jpsi_gluon} we show distributions
of the cross section on the plane $p_t(J/\psi) \times p_t$ 
(matrix element gluon or last gluon in the ladder) for the Kwieci\'nski
UGDF with running scale (left part) and BFKL UGDF (right part). 
Comparing these distributions we conclude that the gluons from the ladder
(LFL - last from the ladder)
contribute to lower transverse momenta than those associated with
the matrix element $ g + g \to J/\psi + g $ (ME) for the Kwieci\'nski UGDF,
where at $p_t$(gluon) $>$ 5 GeV
the matrix element gluons dominate over the ladder gluons.
For the BFKL gluons the situation is much more complicated as here
the distribution for ME gluons and LFL gluons are similar.

\begin{figure}[htb]    % 
 {\includegraphics[width=0.4\textwidth]{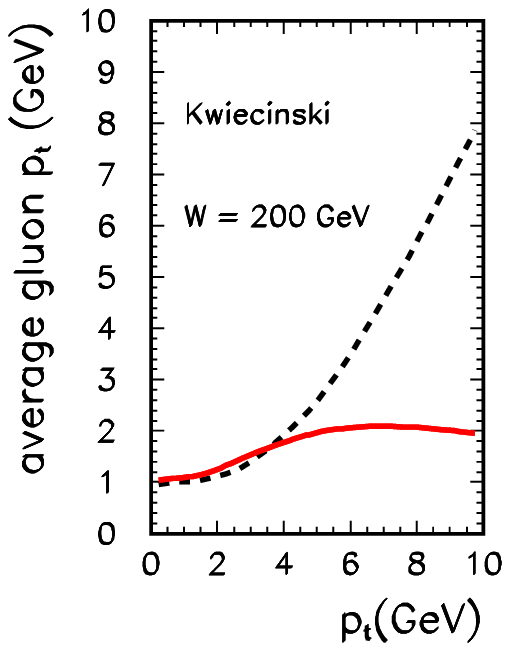}}
 {\includegraphics[width=0.4\textwidth]{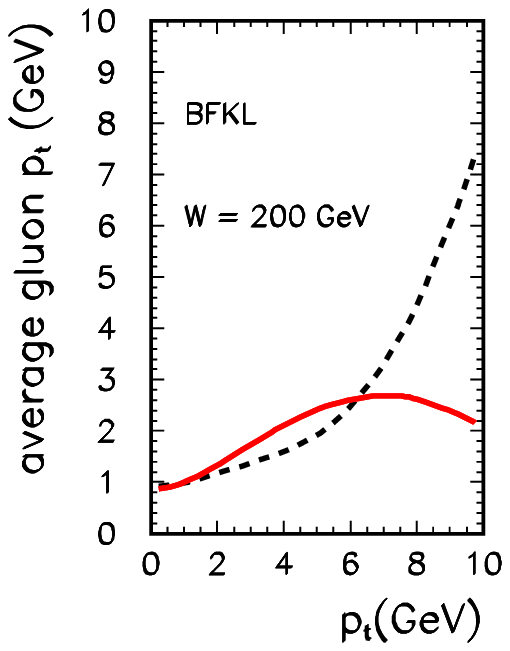}}
\caption{\label{fig:average_ptk1tk2t}
Average transverse momenta of the ME (dashed) and LFL (solid) gluons as
a function of the $J/\psi$ transverse momentum for the Kwieci\'nski UGDF
with running scale (left panel) and the BFKL UGDF (right panel) at
the RHIC energy W = 200 GeV.
}
\end{figure}

In Fig.\ref{fig:average_ptk1tk2t} we show average value of transverse
momentum of the matrix element gluon (dashed line) and of the last gluon
from the ladder (solid line) as a function of $J/\psi$ transverse
momentum.
These average values have completely different dependence on
$p_t(J/\psi)$.
While average value of the LFL gluon transverse momentum is only
weakly dependent on $p_t(J/\psi)$, the average value of the ME gluon
transverse momentum grows monotonically with $p_t(J/\psi)$.
At low $J/\psi$ transverse momenta 
$\langle p_t(LFL) \rangle \sim \langle p_t(ME) \rangle$.
At higher $J/\psi$ transverse momenta
$\langle p_t(LFL) \rangle \; < \; \langle p_t(ME) \rangle$.
For the Kwieci\'nski distribution this happens at smaller transverse
momentum than for the BFKL UGDF.

%The latter are, however, not explicit in the formalism of unintegrated
%gluon distributions. In order to investigate this issue for any model
%of the QCD cascade a Monte Carlo version of the ladder emissions would
%be more useful.

%Up to now we discussed only mechanisms when $J/\psi$ is produced as one
%of many hadrons (both mesons and baryons) in the final state.
%Recently an exclusive process with well defined final state
%was discussed in the literature (see e.g. \cite{exclusive_Jpsi_SS07}
%and references therein). Due to its specificity this final channel is
%not included in and process discussed up to now.
%We wish to discuss now separately a contribution of $J/\psi$ from
%the exclusive production $p p \to p p J/\psi$ due to photon-pomeron or
%pomeron-photon fusion (for details see the analysis in 
%\cite{exclusive_Jpsi_SS07}). 
%In Fig.\ref{fig:dsig_dpt_exclusive} we present invariant distribution
%of $J/\psi$ calculated as in Ref.\cite{exclusive_Jpsi_SS07} for central
%and intermediate rapidity range. This contribution consitutes only a
%small fraction of the inclusive cross section and is concentrated mainly
%at small transverse momenta. The identification of such a component
%would require a measurement of forward/backward protons in addition to
%more centrally produced $J/\psi$. It is not clear to us if this will
%be possible with the future RHIC instrumentation. Such a process is
%currently being analysed at Tevatron \cite{exclusive_Jpsi_experiment}.

\begin{figure}[htb]    % 
 {\includegraphics[width=0.6\textwidth]{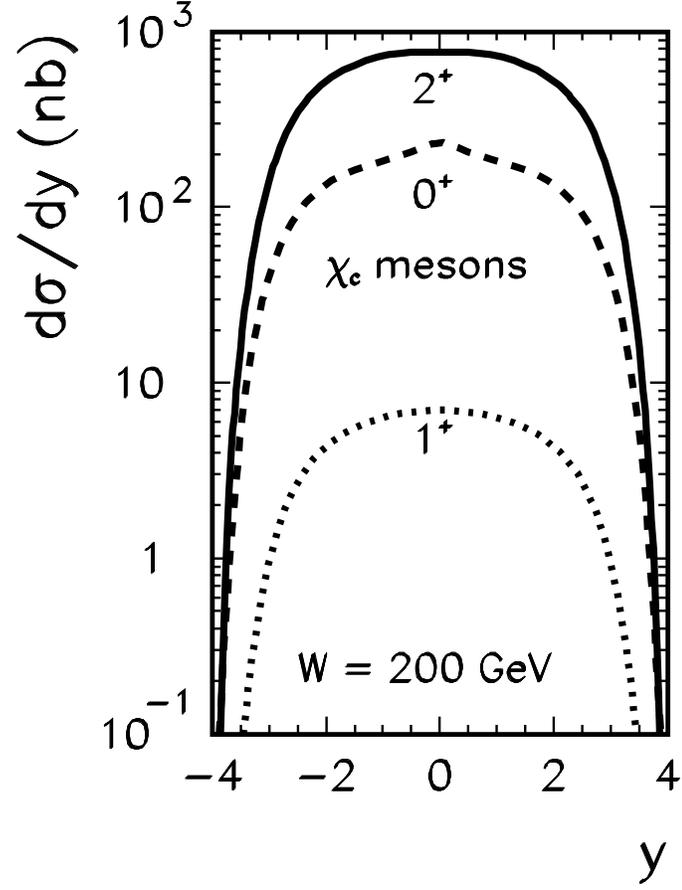}}
   \caption{\label{fig:dsig_dy_chic_predictions}
\small Rapidity distribution of $\chi_c(0^+)$ (dashed), $\chi_c(1^+)$
(dotted) and $\chi_c(2^+)$ (solid) for the RHIC energy obtained with the
Kwieci\'nski UGDF ($b_0$ = 1 GeV$^{-1}$, $\mu^2 = p_t^2(\chi_c)$).
 }
\end{figure}

Our calculations presented up to now show that the production of
$J/\psi$ through radiative decays of $\chi_c$ mesons is one of two
dominant mechanisms. It would be worth to verify this theoretical
prediction experimentally.
This would require measuring the $\chi_c$ mesons independently.
In Fig.\ref{fig:dsig_dy_chic_predictions} we show the distributions
in rapidity for $\chi_c(0^+)$, $\chi_c(1^+)$ and $\chi_c(2^+)$.
These results were obtained with the Kwieci\'nski UGDF,
which seems to be the most reliable for the RHIC energy range. 

\begin{figure}[htb]    % 
 {\includegraphics[width=0.6\textwidth]{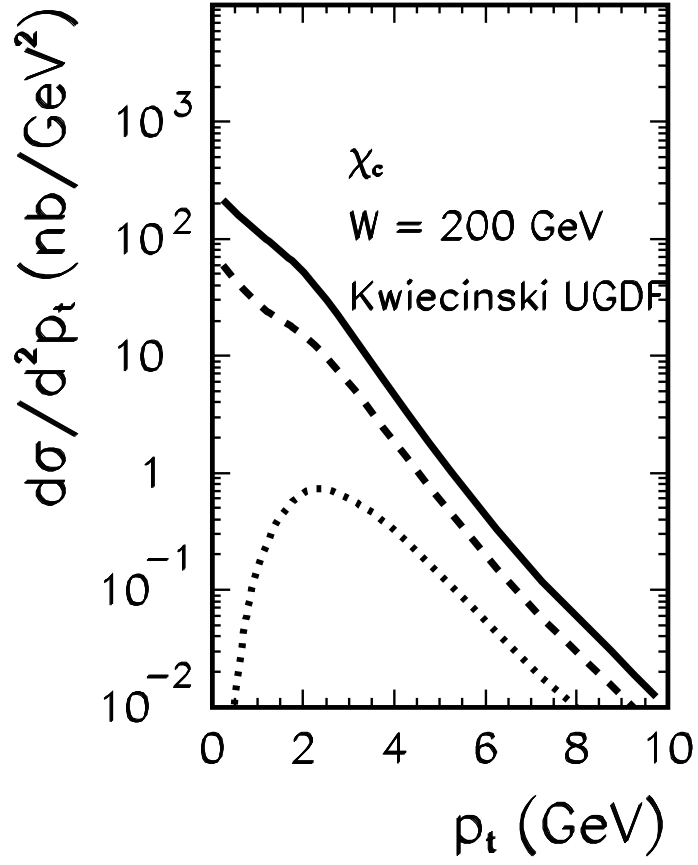}}
   \caption{\label{fig:dsig_dpt_chic_kwiec}
\small 
Transverse momentum distribution of $\chi_c(0^+)$ (dashed), $\chi_c(1^+)$
(dotted) and $\chi_c(2^+)$ (solid) for the RHIC energy obtained with the
Kwieci\'nski UGDF ($b_0$ = 1 GeV$^{-1}$, $\mu^2 = p_t^2(\chi_c)$).
 }
\end{figure}

For completeness in Fig.\ref{fig:dsig_dpt_chic_kwiec} we show the 
corresponding distributions in transverse momentum. In this calculation
-1 $< y <$ 1. 
We wish to point out that the cross sections show no singularity at small
transverse momentum. This contrasts with the collinear factorization
predictions, which are either unphysicsl ($\delta$-like) or divergent
(if based on a $2\to 2$ subprocess $g+g\to\chi_c+g$).
%\marginpar{$\surd$ 14}
There is also a significant difference in shape between the transverse 
momentum distribution for $\chi_c(1^+)$ meson and those for $\chi_c(0^+)$ 
and $\chi_c(2^+)$ mesons. This property emerges from Landay-Yang theorem
which prohibits the coupling of vector states to massless photons (just 
because of quantum numbers incompatible with Bose statistics). 
The production of $\chi_c(1^+)$ states at small $p_T$ is strongly 
suppressed because the initial gluons are almost on shell. 
The suppression goes away at higher $p_T$, as the off-shellness of the 
initial gluons becomes larger. These features are discussed in detail 
in Ref. \cite{Chi_J}.

\begin{figure}[htb]    % 
 {\includegraphics[width=0.6\textwidth]{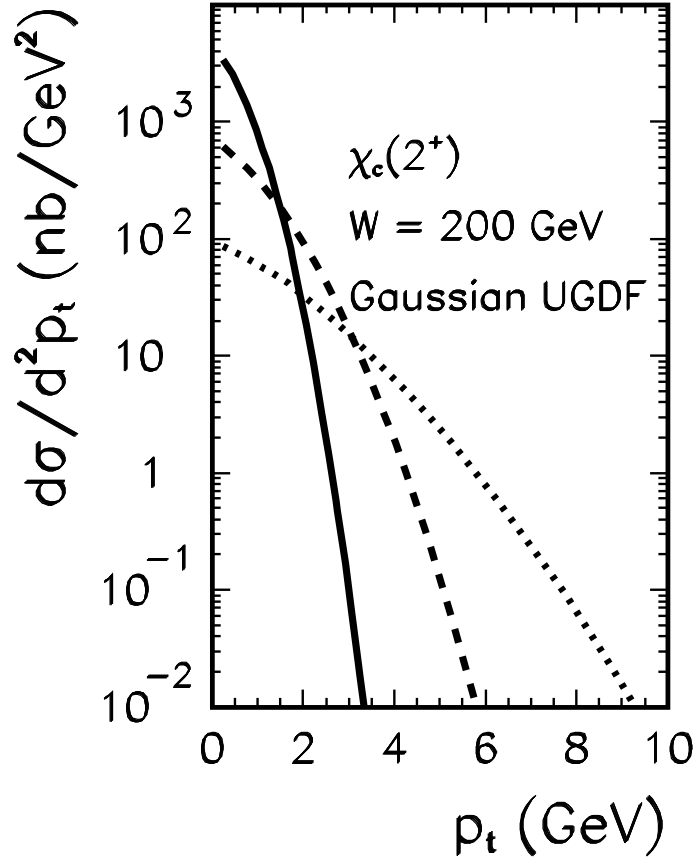}}
   \caption{\label{fig:dsig_dpt_chic2_gauss}
\small 
Transverse momentum distribution of $\chi_c(2^+)$
for the RHIC energy obtained with the Gaussian UGDF and different values
of $\sigma_0$ = 0.5, 1.0, 2.0 GeV.
}
\end{figure}

In contrast to transverse momentum distribution of $J/\psi$ from
the color-singlet mechanism, the distributions of $\chi_c$ mesons
(and consequently the distribution of $J/\psi$ from radiative decays)
strongly depend on the model of UGDF. In particular, in the limiting case
of vanishing initial gluon transverse momenta: $d \sigma/ d^2 p_t \propto
\delta^2 (\vec{p}_t)$. For illustrating the effect quantitatively in
Fig.\ref{fig:dsig_dpt_chic2_gauss} we present transverse momentum
distributions of $\chi_c(2^+)$ for the Gaussian UGDF with different
values of the smearing parameter $\sigma_0$ = 0.5, 1, 2 GeV.
The example clearly demonstrates that a measurement of transverse
momentum distribution of $\chi_c$ mesons would open a new and unique
possibility to test model unintegrated gluon distributions.

%and 
%\ref{fig:dsig_dpt_chic_predictions} we show the predicted cross section
%for $\chi_c(0^+)$, $\chi_c(1^+)$ and $\chi_c(2^+)$ production at RHIC energy
%W = 200 GeV.

In principle the $\chi_c$ mesons (mainly $\chi_c(1^+)$ and $\chi_c(2^+)$)
can be identified via photon-$J/\psi$ decay channel.
At RHIC the $\chi_c$ production mechanism could be also
identified using the $\pi^+ \pi^-$ and $K^+ K^-$ final channels.
The corresponding branching ratios are \cite{PDG}:\\  
%\marginpar{$\surd$ 15}
$BR (\chi_c(0^+) \to \pi^+ \pi^-)$ = 7.2 $\pm$ 0.6 $\times$ 10$^{-3}$,
$BR (\chi_c(0^+) \to K^+ K^-)$ = 5.4 $\pm$ 0.6 $\times$ 10$^{-3}$,\\
$BR (\chi_c(2^+) \to \pi^+ \pi^-)$ = 2.14 $\pm$ 0.25 $\times$ 10$^{-3}$,
$BR (\chi_c(2^+) \to K^+ K^-)$ = 7.7 $\pm$ 1.4 $\times$ 10$^{-4}$.\\
%
%$BR (\chi_c(0^+) \to \pi^+ \pi^-)$ = 7.5 $\pm$ 2.1 $\times$ 10$^{-3}$,
%$BR (\chi_c(0^+) \to K^+ K^-)$ = 7.1 $\pm$ 2.4 $\times$ 10$^{-3}$,\\
%$BR (\chi_c(2^+) \to \pi^+ \pi^-)$ = 1.9 $\pm$ 1.0 $\times$ 10$^{-3}$,
%$BR (\chi_c(2^+) \to K^+ K^-)$ = 1.5 $\pm$ 1.1 $\times$ 10$^{-3}$,\\
%Then the relevant cross section must be multiplied by the corresponding
%branching fraction.

Now we are coming to the issue of \J spin alignment, which was, and still 
%\marginpar{$\surd$ 16}
is under intense debates in the literature. We want to stress once again
that measuring the polarizaton of quarkonium states produced at high 
energies may serve as a crucial test discriminating the different 
concepts of parton dynamics. 

The polarization state of a vector meson is characterized by the spin
alignment parameter $\alpha$ which is defined as a function of any
kinematic variable as
\begin{equation}\label{alpha}
 \alpha(\p)=(d\sigma/d\p -3d\sigma_L/d\p)/(d\sigma/d\p +d\sigma_L/d\p),
\end{equation}
where $\sigma$ is the reaction cross section,
$\p$ is a selected kinematical variable
 and $\sigma_L$ is the part
of cross section corresponding to mesons with longitudinal polarization
(zero helicity state). The limiting values $\alpha=1$ and $\alpha=-1$
refer to the totally transverse and totally longitudinal polarizations.
Here we consider only the behavior of $\alpha$ as a function of the
\J transverse momentum: $\p\equiv |{\mathbf p}_{T}|$.
The experimental definition of $\alpha$ is based on measuring the
angular distributions of the decay leptons
\begin{equation}\label{dgamma}
d\Gamma(\j{\to}\mu^+\mu^-)/d\cos\theta\sim 1+\alpha\cos^2\theta,
\end{equation}
where $\theta$ is the polar angle of the final state muon measured in
the decaying meson rest frame.

\begin{figure}[htb]    % 
 {\includegraphics[width=0.6\textwidth]{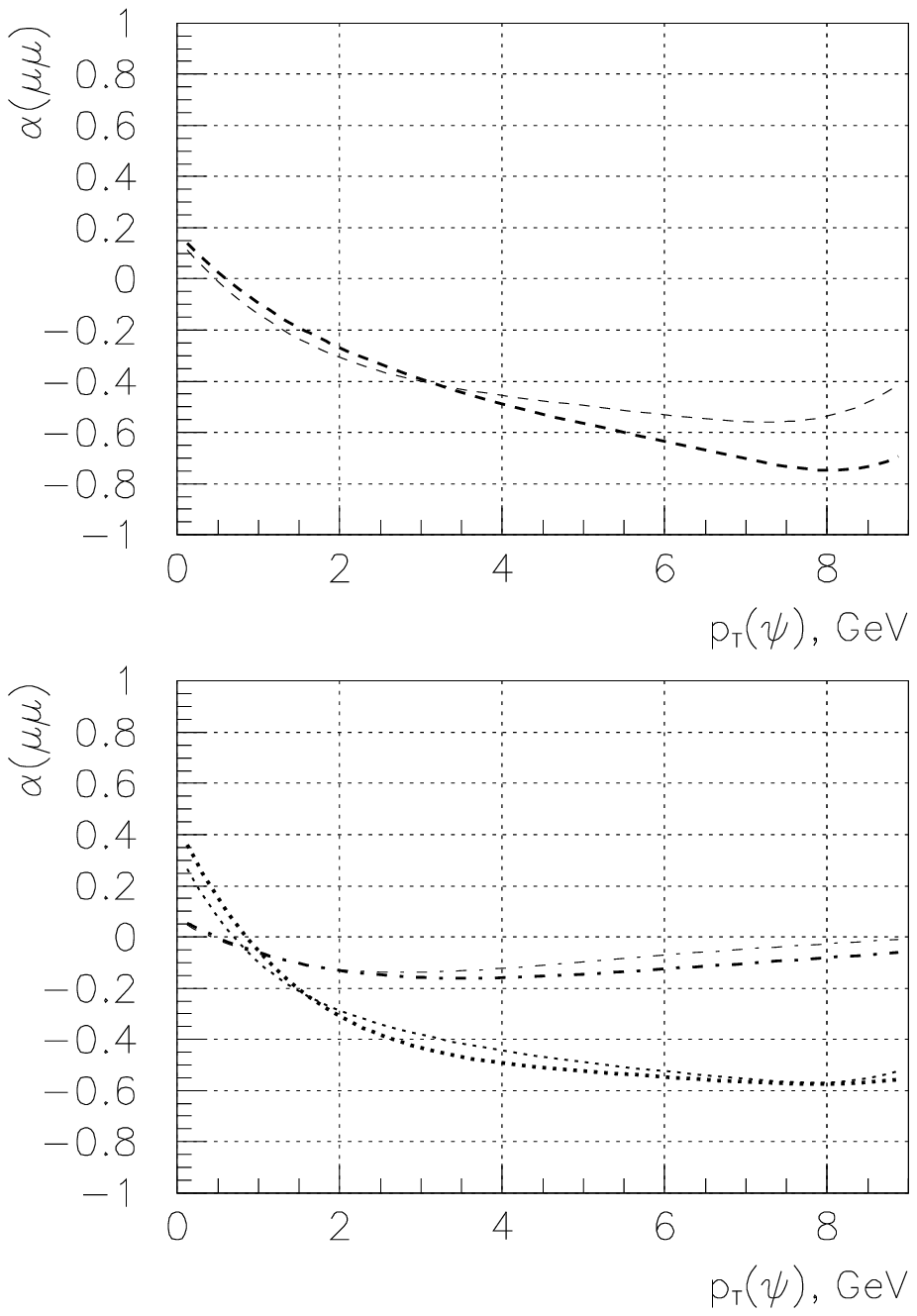}}
   \caption{\label{fig:spin_alignment}
\small
Predictions for the spin-alignment parameter $\alpha$ for \J 
and W = 200 GeV. Thick lines correspond to the Bluemlein parametrization
\cite{Bluem} and the thin lines correspond to the derivative
UGDF parametrization and the GRV collinear distribution.
The top panel is for direct contribution only. The bottom panel
includes the feed-down from $\chi_c$ decays taken into account.
The dotted lines are for the quark spin conservation hypothesis,
and the dash-dotted lines are for the full depolarization hypothesis. 
}
\end{figure}

The results of our calculations for the kinematic conditions of RHIC
are displayed in Fig.\ref{fig:spin_alignment}. 
In order to show the theoretical uncertainty band connected with the 
choice of UGDF, we use two different parametrizations, which are known 
to show the largest difference with each other, namely, the ones proposed 
in Refs. \cite{GLR83} (called 'derivative' for brevity) and
the one from Ref.\cite{Bluem}.

The upper panel in Fig.\ref{fig:spin_alignment} shows the behavior
of the spin alignment parameter $\alpha$ for \J mesons produced in
the direct subprocess (\ref{direct}).
The increase in the fraction of longitudinally polarised mesons comes
from the increasing virtuality (and longitudinal polarization) of the 
initial gluons. These predictions shown here are also valid for $\psi'$ 
mesons.

As far as the contribution from $P$-waves is concerned, nothing is known
on the polarisation properties of their decays. If we assume that the
quark spin is conserved in radiative transitions, and the emission of
a photon only changes the quark orbital momentum (as it is known to be
true in the electric dipole transitions in atomic physics,
$\Delta S=0$, $\Delta L=\pm 1$), then the predictions on $\alpha$
appear to be similar to those made for the direct channel (see lower
panel in Fig.\ref{fig:spin_alignment}, dotted curves). 
If, on the contrary, we assume that the transition 
$\chi_c{\to}J/\psi{+}\g$ leads to a complete depolarization, then we arrive 
at a more moderate behavior of the parameter $\alpha$ (dash-dotted curves 
in Fig.\ref{fig:spin_alignment}). The overall polarization remains
slightly longitudinal ($\alpha\simeq-0.1$) in the whole range of
$p_T$ due to the 'direct' contribution. 
A comparison between the data on \J and $\psi'$ polarization at the
Tevatron \cite{CDF4} seems to give support to the depolarization
hypothesis. The difference between the \J and $\psi'$ polarization
data can be naturally explained by the presence of the depolarizing
contribution in the case of \J and the absence of this contribution
in the case of $\psi'$.

%-----------------------------------
%\section{Figure captions}
%-----------------------------------

%----------------------------------------------------------------------

%-----------------------------------------------------------------------

%------------------------------------------------------------------------

%-----------------------------------------------------------------------

%-----------------------------------------------------------------------

%----------------------------------------------------------------------

%---------------------------------------------------------------------

%---------------------------------------------------------------------

%----------------------------------------------------------------------

%----------------------------------------------------------------------

%----------------------------------------------------------------------

%----------------------------------------------------------------------

%-----------------------------------------------------------------------

%------------------------------------------------------------------------

%------------------------------------------------------------------------

%\begin{figure}[htb]    % 
% {\includegraphics[width=0.4\textwidth]{inv_pt_exclusive_central.eps}}
% {\includegraphics[width=0.4\textwidth]{inv_pt_exclusive_inter.eps}}
%   \caption{\label{fig:dsig_dpt_exclusive}
%\small
%Invariant cross section for the exclusive photoproduction mechanism.
%The left panel is for central rapidity range (-0.35 $< y <$ 0.35) and
%the right panel for intermediate rapidity range (1.2 $< y <$ 2.2).
% }
%\end{figure}

%--------------------------------------------------------------------------

%--------------------------------------------------------------------------

%--------------------------------------------------------------

%--------------------------------------------------------------

%--------------------------------------------------------------

%--------------------------------------------------------
\section{Discussion and Conclusions}
%--------------------------------------------------------

%\marginpar{$\surd$ 17}
We have considered different mechanisms contributing to the inclusive
production of \J mesons in $pp$ collisions at RHIC kinematics. The
outcome of our study  is the following.

We have inspected the hierarchy of contributions and found that the
dominant contribution to the cross section comes from radiative 
decay of $\chi_c$ mesons, mainly from $\chi_c(2^{++})$ state. 
The second most important mechanism is the direct color-singlet production.
The sequential process through the intermediate $\psi'$ turned out to be
nonnegligible and constitutes about a quarter of the direct color-singlet
contribution. To our knowledge, these processes were not included in
previous calculations in the literature on the subject.

As a by-product, we have demonstrated the advantage of the \ktf approach
in calculating the $\chi_c$ spectra: the latter can hardly be calculated
in a cocsistent way in the collinear scheme. In order to verify the
production mechanism suggested in our analysis, we have proposed an
independent measurement of inclusive $\chi_c$ cross sections in the
$\pi^+\pi^-$ and $K^+K^-$ decay channels.

We have applied our approach to describe the data on inclusive \J
production recently collected by the PHENIX Collaboration at the BNL.
Both rapidity and transverse momentum distributions have been discussed.
The new precise data at small \J transverse momenta appeared to show
very strong analysing power, imposing stringent constraints on
%in selecting the 
unintegrated gluon distributions. The best description of the data
is obtained with the UGDF proposed by Kwieci\'nski.

Another piece of important information on the underlying gluon dynamics
can be extracted from studying kinematic correlations between \J mesons
and coproduced gluon jets. In this paper we have presented our predictions
for the two dominant contributing mechanisms. 

%We have found that discussed recently in the literature exclusive
%production of $J/\psi$ due to pomeron-photon or photon-pomeron fusion
%constitutes only small fraction of the inclusive cross section.

%We do not find much room for color-octet contribution which was advocated
%for some time as a solution to the $J/\psi$ puzzle.

Finally, we have presented our predictions on the \J spin alignment.
The latter can serve as important test discriminating two different
concepts of parton model.

In the present paper we have discussed mechanisms of $J/\psi$ production 
in elementary collisions. We believe that our findings here may be also 
useful for nuclear collisions, where $J/\psi$ suppression was originally
suggested as a useful indication of the presence of the quark-gluon 
plasma.

\vskip 0.5cm

{\bf Acknowledgments} 
We are indebted to Abigail Bickley from the PHENIX collaboration
for providing us recent experimental data on inclusive $J/\psi$
production measured at RHIC. 
Thanks go to Jerzy Bartke for pointing to us the possibility
that the $\psi'$-decay contribution may be sizeable.
%We are also indebted to Tomasz Pietrycki for help in preparing some 
%more complicated figures.
%The discussion with Wolfgang Sch\"afer is greatly acknowledged.
This work was partially supported by the grant
of the Polish Ministry of Scientific Research and Information Technology
number 1 P03B 028 28.

%FIGURES

\newpage

%======================================================================

\end{document}